\documentclass[fleqn,usenatbib]{mnras}

\usepackage{newtxtext,newtxmath}
\usepackage[T1]{fontenc}
\usepackage{ae,aecompl}

\usepackage{graphicx}	% Including figure files
\usepackage{amsmath}	% Advanced maths commands
\usepackage{amssymb}	% Extra maths symbols
\usepackage{subfigure}
\title[Chemical evolution of UFDs: testing the IGIMF]{Chemical evolution of ultra-faint dwarf galaxies: testing the IGIMF}

\author[E. Lacchin et al.]{
E. Lacchin,$^{1,2,3}$\thanks{E-mail: elena.lacchin2@unibo.it}
F. Matteucci,$^{3,4,5}$\thanks{E-mail: matteucci@oats.inaf.it}
F. Vincenzo$^{6}$
M. Palla$^{3,7}$
\\
$^{1}$INAF - OAS, Osservatorio di Astrofisica e Scienza dello Spazio di Bologna, via Gobetti 93/3, I-40129 Bologna, Italy\\
$^{2}$Department of Physics and Astronomy, University of Bologna, via Gobetti 93/3, 40129 Bologna, Italy\\
$^{3}$Dipartimento di Fisica, Sezione di Astronomia, Universit\'{a} degli Studi di Trieste, via G. B. Tiepolo 11, I-34131, Trieste, Italy\\
$^{4}$INAF, Osservatorio Astronomico di Trieste, via G. B. Tiepolo 11, I-34131, Trieste, Italy\\
$^{5}$INFN, Sezione di Trieste, via A. Valerio 2, I-34100, Trieste, Italy\\
$^{6}$Center for Cosmology and AstroParticle Physics, The Ohio State University, 191 West Woodruff Avenue, Columbus, OH 43210, USA\\
$^{7}$ IFPU - Institute for Fundamental Physics of the Universe, Via Beirut 2, I-34014, Trieste, Italy\\
}

\date{Accepted XXX. Received YYY; in original form ZZZ}

\pubyear{2020}

\begin{document}
\label{firstpage}
\pagerange{\pageref{firstpage}--\pageref{lastpage}}
\maketitle

\defcitealias{recchi}{R14}

\begin{abstract}

We test the integrated galactic initial mass function (IGIMF) on the chemical evolution of 16 ultra-faint dwarf (UFD) galaxies discussing in detail the results obtained for three of them: Bo\"otes I, Bo\"otes II and Canes Venatici I, taken as prototypes of the smallest and the largest UFDs. These objects have very small stellar masses ($\sim 10^3-10^4 \mathrm{M_{\odot}}$) and quite low metallicities ([Fe/H]$<-1.0$ dex). We consider four observational constraints: the present-day stellar mass, the [$\alpha$/Fe] vs. [Fe/H] relation, the stellar metallicity distribution function and the cumulative star formation history. Our model follows in detail the evolution of several chemical species (H, He, $\alpha$-elements and Fe). We take into account detailed nucleosynthesis and gas flows (in and out). Our results show that the IGIMF, coupled with the very low star formation rate predicted by the model for these galaxies ($\sim 10^{-4}-10^{-6}\ \mathrm{M_{\odot}yr^{-1}}$), cannot reproduce the main chemical properties, because it implies a negligible number of core-collapse SNe and even Type Ia SNe, the most important polluters of galaxies. On the other hand, a constant classical Salpeter IMF gives the best agreement with data, but we cannot exclude that other formulations of the IGIMF could reproduce the properties of these galaxies. Comparing with Galaxy data we suggest that UFDs could not be the building blocks of the entire Galactic halo, although more data are necessary to draw firmer conclusions.

\end{abstract}


%%%%%%%%%%%%%%%%%%%%%%%%%%%%%%%%%%%%%%%%%%%%%%%%%%

%%%%%%%%%%%%%%%%% BODY OF PAPER %%%%%%%%%%%%%%%%%%

\section{Introduction}

Over the last decades, a large number of satellite galaxies have been found orbiting the Milky Way. They are characterized by low surface brightness and small effective radius making them difficult to be detected. With the exception of the two Magellanic Clouds, which have been known since the ancient times, after the discovery of Sculptor \citep{shapley1938} and until 2005, only eight other satellites were detected and were named classical Dwarf Spheroidal galaxies (dSphs) for their small dimensions. With the development of digital surveys, like the Sloan Digital Sky Survey (SDSS), new and fainter galaxies were discovered which were classified as Ultra-Faint Dwarf galaxies (UFDs). No formal distinction between UFDs and dSphs can be found in literature even though most of the studies fix the separation between $M_{V}=-7.7$ and $-8.0$ mag \citep{simonegeha2007,simon2019}: dwarf galaxies fainter than these values are classified as UFDs while the others as dSphs. 
UFDs are considered the most dark matter dominated systems observed today in the Universe, thus they are studied in order to constrain the nature of the dark matter \citep{spekkens2013,Ackermann2014,kennedy2014,
regis2017,jeltema2016,brandt2016,penarrubia2016}. 
From color-magnitude diagram (CMD) fitting analysis it emerges that UFDs host very old stellar populations (\ $\gtrsim  10-12\ \mathrm{Gyr}$; \citealp{okamoto2012}) while from spectroscopic studies it has been found that most of the stars are very (VMP, [Fe/H] $<-2.0$ dex from \citealp{beers2005}) to extremely (EMP, [Fe/H] $<-3.0$ dex) metal-poor ones. All such peculiar features make the UFDs a perfect environment to understand how the nucleosythesis proceeded in the early Universe and verify whether a first generation of very massive and metal-free stars (so-called Population III) might have existed \citep{salvadorieferrara2009}. %These stars should have enriched the ISM with some metals, leaving a signature in the subsequent stellar generations, hosted now in UFDs \citep{salvadori2019}. 

%The enhancement in carbon observed in almost all UFDs is one of them: Carbon-Enhanced Metal-Poor (CEMP) stars, which display [C/Fe] $\geqslant$ +0.7 dex \citep{aoki2007} and represent a consistent fraction of stars at very low metallicities, could have been imprinted by an earlier generation of stars \citep{salvadori2015,debennassuti2014b}. 
%In addition, studying a C-enhanced damped Ly$\alpha$ absorption system (DLA), called also PopIII DLAs, \citet{salvadorieferrara2012} derived that these objects could be the gas-rich counterparts of the smallest dwarf galaxies. However, \citet{skuladottir2018} pointed out that DLA star formation histories are different from the one of Scuptor dSph.

%Studying a C-enhanced damped Ly$\alpha$ absorption system \cite{salvadorieferrara2012} pointed out that Carbon-Enhanced Metal-Poor (CEMP) stars, which display [C/Fe] $\geqslant$ +0.7 dex \citep{aoki2007} and represent a consistent fraction of stars at very low metallicities, could have been imprinted by an earlier generation of stars. 

Moreover, the cosmological $\Lambda$ cold dark matter ($\Lambda$CDM) paradigm predicts that the large structures observed today in the Universe are the result of the merging of small systems in increasingly larger dark matter halos. In this scenario, dSphs have been proposed to be the survived progenitors of the halo component of the Galaxy \citep{helmi1999,bullock2001,harding2001,bullock2005,delucia2008}. However, this hypothesis faded when deeper analyses on dSphs have been carried on \citep{helmi2006, catelan2009,fiorentino2015} showing that the dSphs have a different abundance patterns from halo stars. With the discovery of a large number of smaller satellite galaxies, the interest has been shifted on these systems, the UFDs. \cite{spitoni2016} modeled the chemical evolution of the Galactic halo both assuming it to be formed from the accretion of disrupted satellites as well as from the infall of pre-enriched gas. They ruled out the possibility that the Galactic halo was entirely originated by the merging of current dSphs and UFDs ancestors. However, they do not exclude that dwarf galaxies provided a contribution to the halo formation. %They    makes  that to establish whether these objects could have been the relics of the first galaxies and the building blocks of the Galactic halo, as suggested by the hierarchical clustering scenario of galaxy formation.

The aim of this work is to study the chemical evolution of the gas and its chemical abundances in the interstellar medium of sixteen UFD galaxies starting from the available observational constraints: the chemical abundances derived today in the atmosphere of their stars and the present-day stellar masses (if not available, we have used the visual magnitudes) and the cumulative SFH, if available. 
In particular, we have focused our attention on the effects of the Initial Mass Function (IMF), one of the most important ingredients to derive the chemical enrichment history and the stellar abundance patterns of a galaxy together with the star formation rate (SFR). The IMF represents the mass distribution function of stars at their birth and its most widespread parametrization is the Salpeter IMF \citep{Salpeter1955}, derived in the solar vicinity. Nowadays, it is not clear whether the IMF is a universal function or it depends on the environment, nor if it is constant in time \citep{kroupa2002,ferreras2016}. Recently, a more detailed formulation of the IMF was proposed by \citet{kroeweid2003} and  \citet{weidekro2005}, the so-called Integrated Galactic Initial Mass Function (IGIMF). Generally, the IGIMF depends on the star formation rate and, in some parametrizations, also on metallicity, thus it represents a more physical formulation than the canonical Salpeter IMF. For this reason, it is very important to test the IGIMF in peculiar environments different from the solar neighborhood.% \cite{recchi} tested it on ellipticals while \cite{vincenzo2015} on dwarf spheroidal galaxies (dSphs).

Therefore, in this work, we test, for the first time in literature, the IGIMF on the evolution of UFDs. In particular, we adopt the mild model proposed by \citet[hereafter \citetalias{recchi}]{recchi}, which depends upon the SFR and the metallicity, in a detailed chemical evolution model. Such a model is based on the work of \cite{lanfranchiematteucci2004} and follows the evolution of the gas abundances of many chemical elements, from lighter (H, D, He, Li) to heavier (C, N, $\alpha$-elements, Fe-peak elements, s- and r-process elements). %We take into account detailed stellar nucleosynthesis, supernova (Ia, Ib, Ic, II) progenitors as well as gas infall and outflow.
The same model has been later adopted by \cite{lanfranchi2006a,lanfranchi2006b,lanfranchi2008}, \cite{lanfranchi2007}, \cite{cescutti2008}, \cite{lanfranchi2010}, \cite{vincenzo2014}, \cite{vincenzo2015}. %, \cite{}, \.
In particular, \cite{vincenzo2014}, studied the chemical evolution of two UFDs (Bo\"otes I and Hercules, which are also studied in this work) by assuming the Salpeter IMF. They concluded that UFDs are characterized by extremely low star formation efficiencies (SFEs, $\nu=0.001-0.01\mathrm{Gyr^{-1}}$) as it has been pointed out by \citet{salvadorieferrara2009}, even lower than the one found for dSphs ($\nu=0.1\mathrm{Gyr^{-1}}$). The main effect of low SFEs is the lower [Fe/H] at which Type Ia SNe start polluting the ISM in the plots [$\alpha$/Fe] vs. [Fe/H]. This is a consequence of the time-delay model for the chemical enrichment as pointed out in \citet{lanfranchiematteucci2004}. The same numerical code used by \cite{vincenzo2014} was earlier adopted also by \cite{koch2013} to study the chemical evolution of the Hercules UFD. All these works about UFDs derived also a very short time-scale ($\tau_{inf}=0.005$ Gyr) for the accretion of gas in the DM halos.

Finally, \cite{vincenzo2015} tested the IGIMF of \citetalias{recchi} in the chemical evolution of Sagittarius dSph galaxy concluding that the IGIMF better reproduces the observed [$\alpha$/Fe], [Eu/Fe] and [explosive-to-hydrostatic] $\alpha$-element ratios in this galaxy than the Salpeter and the Chabrier IMFs \citep{chabrier2003}. 
The effects of the IGIMF have also been tested in the chemical evolution of the solar neighborhood \citep{calura2010}, local elliptical galaxies \citep{recchi2009,demasi2018} and in high-redshift starbursts \citep{palla2019}. 

The paper is organized as follows: in Section \ref{sec:igimf} we describe the IGIMF theory while in Section \ref{sec:chemmodel} we focus on the adopted chemical evolution model. The observational data are summarized in Section \ref{sec:obsdata} and in Section \ref{sec:results} the results we have obtained are presented. Finally, in Section \ref{sec:conclusions} some conclusions are drawn.

\section{The integrated galactic initial mass function}
\label{sec:igimf}

From stellar counts, Salpeter derived a one-slope IMF expressed as:

\begin{equation}
\varphi(m)=Am^{-(1+x)},
\end{equation}
where $x=1.35$ and A is the normalization factor derived by imposing:
\begin{equation}
\int_{0.1M_{\odot}}^{100M_{\odot}} m\varphi(m) dm=1.
\end{equation}
This IMF parametrization is constant in time, thus it is assumed to have the same shape during all the galaxy evolution. The recently proposed IGIMF instead depends on the features of the environment, making it varying with time.

The IGIMF theory is based on the assumption that most stars in a galaxy form in star clusters; this statement was derived from observations of star-forming regions in the Milky Way and led to the conclusion that $70$ to $90\%$ of stars were formed in embedded clusters \citep{lada}.
The remaining stars are supposed to have originated in short-lived clusters that dissolved rapidly. 

For this reason we should introduce the mass distribution function of the embedded clusters, $\xi_{ecl}$, that weights the classical IMF, $\varphi(m)$:
\begin{equation}
\xi_{IGIMF}\ (m,\psi(t))=\int_{M_{ecl,min}}^{M_{ecl,max}(\psi(t))} dM_{ecl}\ \xi_{ecl}(M_{ecl})\ \varphi(m \leq m_{max})
\end{equation}
normalized in mass such that:

\begin{equation}
\int_{m_{min}}^{m_{max}} dm\ m\ \xi_{IGIMF}\ (m,\psi(t))=1.
\end{equation}

In the present work, the IGIMF which has been tested depends both on the SFR and the [Fe/H] value of the gas in the parent galaxy. We have followed the mild model of \citetalias{recchi} based on the following assumptions, derived by observations:
\begin{itemize}

\item the mass distribution function of the embedded clusters is assumed to be a power law of the form, $\xi_{ecl}(M_{ecl}) \propto M_{ecl}^{-\beta}$, where $\beta = 2$ \citep{zhang1999}.
Its lower and upper limits are assumed to be $M_{ecl,min}=5\mathrm{M_{\odot}}$, i.e. the mass of the Taurus-Auriga aggregate, which is the smallest star-forming stellar cluster known, while  \citet{weidner2004}
obtained for $M_{ecl,max}$ a dependence on the SFR:

\begin{equation}
log M_{ecl,max}= A+B\ log\ \psi(t)
\end{equation}
where $A=4.83$ and $B=0.75$.

\item Within each embedded cluster, the stellar IMF is assumed to be invariant and, for this model, the two-slope power law one has been chosen (a simplified version of the multi-slope one used by \cite{weidekro2005} in their original work), which is defined as:

  \begin{equation} 
\varphi(m)= k \begin{cases}
  \ \left(\frac{m}{m_H}\right)^{-\alpha_1} \qquad & \text{, $m_{low} \leq m < m_H$} \\

  \ \left(\frac{m}{m_H}\right)^{-\alpha_2}\qquad  & \text{, $m_H \leq m <m_{max}$}
\end{cases}
\label{eq:stellarimf}
\end{equation}
with exponents:

\begin{align}
\alpha_1= &\ 1.30 \qquad \qquad \qquad \quad \qquad,0.08 \mathrm{M_{\odot}} \leq m <0.5 \mathrm{M_{\odot}}\\
\alpha_2=&\ 2.3+0.0572\cdot \mathrm{[Fe/H]}  \qquad ,0.5 \mathrm{M_{\odot}} \leq m < m_{max}
\label{eq:alpha2}
\end{align}

The novelty introduced by \citetalias{recchi} is the dependence of $\alpha_2$ on metallicity, expressed in the form of [Fe/H] value. This dependence is based on the results of \citet{marks} who, studying the mass distribution of globular clusters in the Milky Way, deduced that in such structures the lower the cluster metallicity, the more top-heavy the IMF.
In particular, they obtained Equation \ref{eq:alpha2} assuming a constant cluster density.

In the mild model used here, metallicity influences only the slope of the IMF in the high-mass range and, as a consequence, the maximum stellar mass $m_{max}$, a quantity that depends also on the SFR.
$m_{max}$, indeed, is a function of the mass of the embedded cluster $M_{ecl}$ since, for low SFRs, the small clusters do not have enough mass to produce very massive stars, while, for large SFRs, the maximum mass an embedded cluster can achieve is very high, thus very massive stars can be formed. %; neverthless it can not overcome the empirical limit which is assumed to be $m_{max_*}=150\ M_{\odot}$ \citep{wek2004}.
\end{itemize}

 \begin{figure} 
 %centering
 \hspace{-0.1cm}
  \includegraphics[width=1.1\columnwidth]{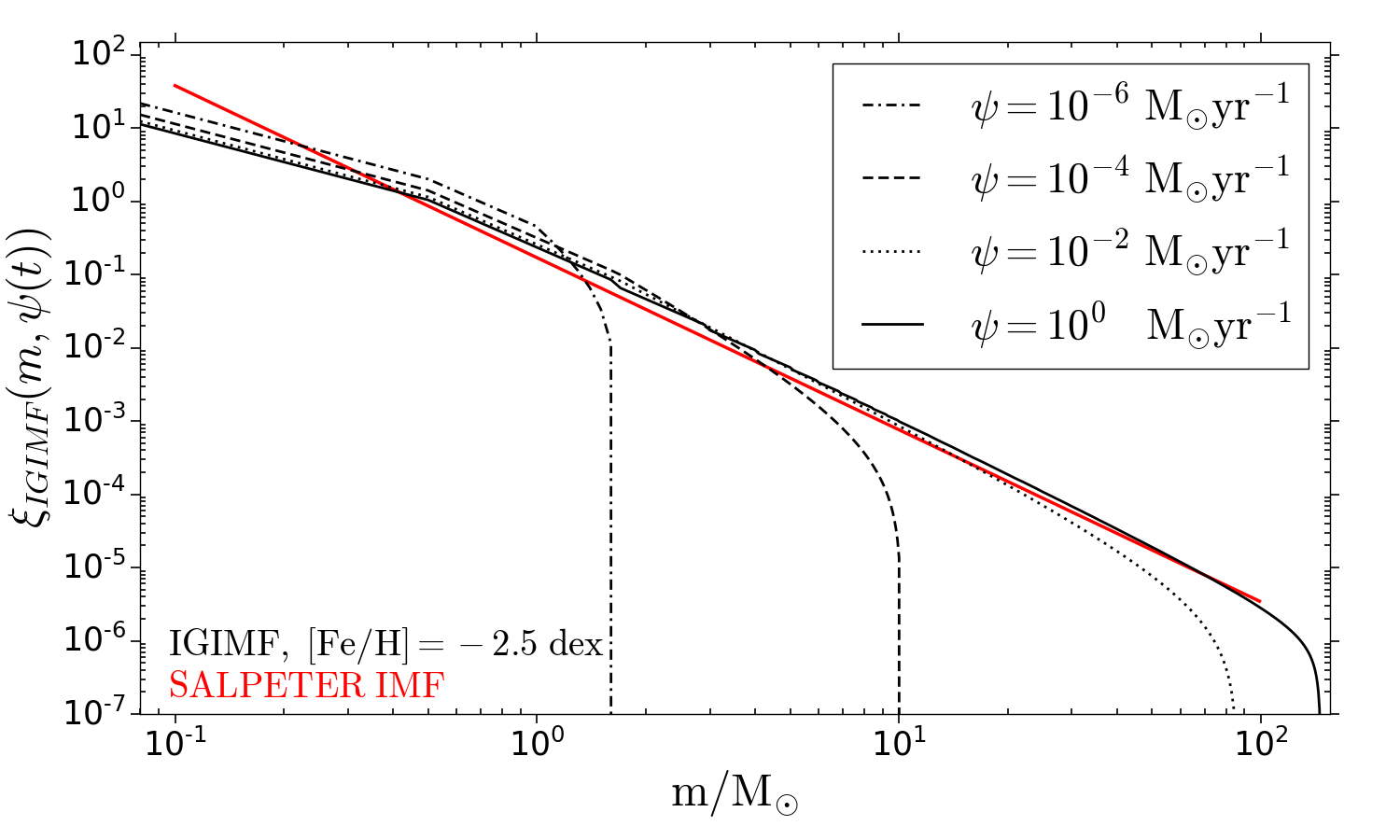}
  \caption{The predicted IGIMF plotted as a function of the stellar mass $m$, obtained for four values of SFR while maintaining fixed the [Fe/H] value. The SFR dependence changes the position of the IGIMF truncation: in particular a decrease of the SFR leads to a shift of the truncation towards lower stellar masses.}
  \label{igimf1}
\end{figure}

\begin{figure} 
 %\centering
 \hspace{-0.1cm}
  \includegraphics[width=1.1\columnwidth]{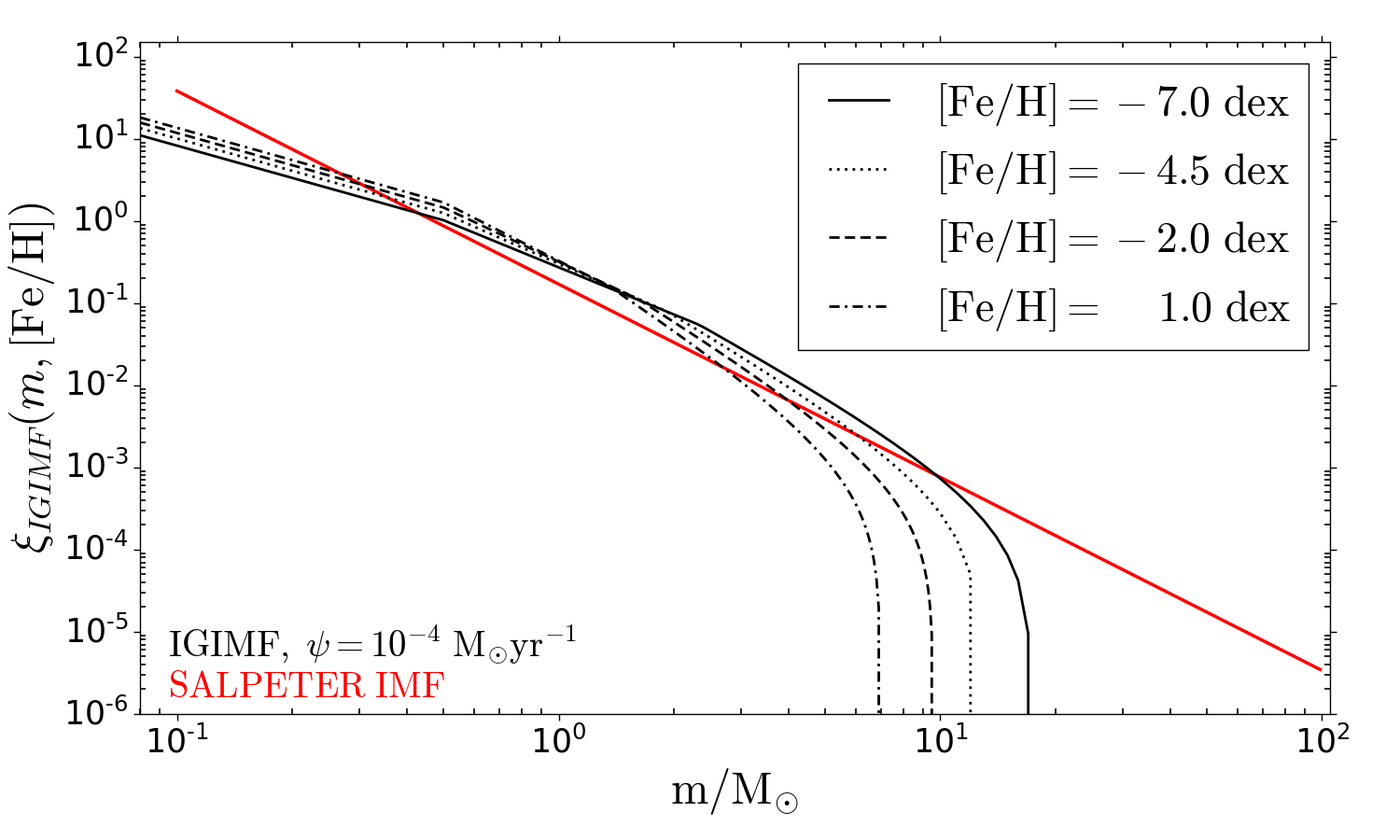}
  \caption{The predicted IGIMF plotted as a function of the stellar mass $m$, obtained for four values of [Fe/H] while maintaining fixed the SFR. The [Fe/H] dependence changes the position of the IGIMF truncation: in particular an increase of the [Fe/H] leads to a shift of the truncation towards lower stellar masses.}
  \label{igimf2}
\end{figure}

In Figures \ref{igimf1} and \ref{igimf2} is shown the IGIMF as a function of the stellar mass compared with the Salpeter IMF.
In Figure \ref{igimf1} the [Fe/H] value is fixed while the SFR is varied; in Figure \ref{igimf2} the [Fe/H] value is varied while the SFR is maintained fixed. What can be inferred is that the IGIMF varies more with the SFR than with the [Fe/H] value, and that the two dependences are opposite (decreasing the SFR leads to a higher truncation, the same that happens increasing the [Fe/H] value).

\section{Chemical evolution model}
\label{sec:chemmodel}
The method used to study the formation and the evolution of UFD galaxies is the same reported in \cite{lanfranchiematteucci2004} where they firstly studied the evolution of dSphs.
Similarly to dSphs, UFDs have been supposed to be formed by the accretion of primordial gas in a pre-existing dark matter (DM) halo, but on smaller time-scales because of their lower mass and radial extension.

The chemical evolution models used in this work permit to follow the evolution of the chemical abundances of several elements\footnote{The models used in this work follow the evolution of the chemical abundances of H, He, C, O, N, Ne, Mg, Si, S, Ca, Fe, Ba, Eu, La, Sr, Y, Zr, Zn, Ni, K, Sc, Ti, Va, Cr, Mn, Co.}.
The main features of the model are:
\begin{itemize}
\item each galaxy is treated as one zone with istantaneous mixing of the gas within it;
\item no istantaneous recycling approximation (IRA) is assumed, so the stellar lifetimes are considered;
\item each galaxy is treated as an open box, thus gas infall and galactic winds are included;
\item the nucleosythesis prescriptions include the  metallicity-dependent stellar yields of \cite{Karakas} for low and intermediate-mass stars while, for massive stars (SNe II and Hypernovae), the ones of \cite{kob2006} are adopted. For SNe Ia we have adopted the yields obtained by the W7 model of \cite{iwamoto}. 
\item we have adopted the single-degenerate scenario for Type Ia SNe progenitors where a C-O white dwarf accretes mass from its red giant companion until its mass reaches the Chandrasekar one ($M_{Ch}=\mathrm{1.44\ M_{\odot}} $) and explodes via C-deflagration \citep{matteuccierecchi2001}. Our formulation of the SNe Ia rate gives very similar results to the ones obtained with a double-degenerate model for SNe Ia (see \citealp{matteucci2009}).
\end{itemize}

\subsection{Basic equations}
The temporal evolution of the gas mass in the form of element $i$ within the ISM is described by the following equation:
\begin{equation}
\dot{M}_{gas,i}(t)=-\psi(t)X_i(t)+R_i(t)+(\dot{M}_{gas,i})_{inf}-(\dot{M}_{gas,i})_{wind}
\label{eq:chemevol}
\end{equation}
where $X_i(t)=M_{gas,i}(t)/M_{gas}(t)$
is the abundance by mass of a generic element $i$, with $\sum_i X_i =1$ where $i$ runs over all the elements that form the gas of the ISM. 

\begin{itemize} 
\item The first term in the right-hand side represents the rate at which the gas mass in the form of element $i$ is subtracted to form stars. The function $\psi(t)$ represents the SFR, i.e. the amount of gas that is turned into stars per unit time which is assumed to follow the Schmidt law with $k=1$ \citep{schmidt1959}:

\begin{equation}
\psi(t)=\frac{\mathrm{d} M_{gas}}{\mathrm{d} t}=\nu M_{gas}^k(t)
\end{equation}

where $\nu$ is the SFE, expressed in terms of $ \mathrm{Gyr^{-1}}$, which is defined as the inverse of the star formation time-scale, the time needed to convert all the gas into stars.

The star formation is assumed to continue after the onset of the galactic wind but at a lower rate, since a fraction of the gas, from there on, is carried out from the galaxy.

\medskip
\item The second term, $R_i(t)$, concerns the restored mass in the form of element $i$ that the stars eject in the ISM per unit time. This term contains all the prescriptions about the stellar yields for low-intermediate mass stars (LIMS) and core-collapse (CC) SNe (Type II, Ib/c) as well as supernova progenitor models (for more details see \citealp{matteucci2001} and \citealp{lanfranchiematteucci2004}).

\medskip
\item The third term regards the gas mass of element $i$ that is accreted during the infall event where for the rate of gas infall is assumed:
\begin{equation}
(\dot{M}_{gas,i})_{inf}\propto X_{i,inf}\cdot e^{-t/\tau_{inf}},
\end{equation}
with $\tau_{inf}$ the infall time-scale, namely the time-scale of mass accretion. As aforementioned, the gas out of which galaxies are assumed to be formed has primordial composition, thus $X_{i,inf}=0$ for all the elements except for hydrogen, deuterium, 3-helium and 4-helium and lithium.

\medskip
 \item The fourth term represents the gas mass of element $i$ that is lost because of galactic wind per unit time.  The rate of gas loss at time $t$ is assumed to be proportional to the total mass of the gas as follows:

\begin{equation}
(\dot{M}_{gas,i})_{wind}=\omega_i \ M_{gas},
\end{equation} 

where $\omega_i$ is a free parameter representing the efficiency of the galactic wind (i.e. mass loading factor, expressed in $\mathrm{Gyr^{-1}}$), as defined in \citet{yin2011}. In this work we have assumed a \textit{normal} wind, thus the efficiencies are equal for every chemical element. $\omega$ contains all the information about the energy released by SNe and stellar winds, as well as the efficiency with which such energy is converted into the gas escape velocity. In particular, the galactic wind is assumed to develop when the thermal energy of the gas, associated with the stellar feedback, becomes larger than its binding energy, which mainly depends on the mass of the dark matter halo (for more details see \citealp{bradamante1998}).
For the stellar feedback we have assumed the same as in \cite{yin2011}. We have also tested the case in which the wind efficiency depends on the SN rates, as done by \citet{romano05}, since for some models with the IGIMF a negligible number of SNe is predicted to explode. This peculiarity is more evident in Bo\"otes II and therefore we will discuss the results we have obtained only for this UFD ( see Section \ref{subsec:boo2}).

% For the stellar feedback prescriptions we have assumed the same as in \cite{yin2011} where the energy released in a single supernova event is equal to $10^{51}$ erg whereas for the stellar wind $E_{sw}=10^{49}$ erg. The efficiencies with which such an energy is used to thermalize the ISM are assumed to be $\eta_{SNIa}=0.8$, $ \eta_{SNII}=0.03$ and $ \eta_{sw}=0.03$.
%The difference between Type Ia efficiency and the other two values is explained by the different conditions of the medium when SNe Ia explosions take place. The medium, at that time, has become hotter and more rarefied thanks to the energy previously released by Type II SNe and stellar winds from massive stars (see \citealp{recchi2001}). In any case, our feedback prescriptions correspond to an average global energy factor of $\sim 30\%$.

\end{itemize}

\section{Data sample}
\label{sec:obsdata}

We have modeled the chemical evolution of 16 UFD galaxies, the only ones with available high-resolution spectroscopic data which are required to compare the model predictions with observations. These galaxies are: Bo\"otes I (Boo I), Bo\"otes II (Boo II), Canes Venatici I (CVn I), Canes Venatici II (CVn II), Coma Berenices (Com), Grus I (Gru I), Hercules (Her), Horologium I (Hor I), Leo IV, Reticulum II (Ret II), Segue I (Seg I), Segue II (Seg II), Triangulum II (Tri II), Tucana II (Tuc II), Tucana III (Tuc III) and Ursa Major II (UMa II).
 
 Here we report only the data samples we have adopted for three of the sixteen UFDs: Boo I, Boo II and CVn I, whose analysis is presented in the next section. We have chosen to discuss only these galaxies since the major difference in our results are related to the mass of the galaxy. We have selected Boo I to represent the most massive UFDs, while Boo II stands for the least massive ones as it is quantified in Table \ref{tab:ufdsfetaures}, where the main observational features are summarized. We have added CVn I to our analysis given its larger mass and spatial extension which make it more similar to a dSph galaxy.

In general, we have chosen, if available, high-resolution data for the study of the [$\alpha$/Fe] vs. [Fe/H] relations, while for the metallicity distribution function (MDF) we have added also the low and medium resolution data to our analysis. Given that not all the observational papers report the abundance values relative to the same solar composition, we have rescaled all of them to the solar photosphere adundances of \cite{asplund2009}, since they are the ones adopted in our chemical evolution models.
Moreover, at the very low metallicities, typical of UFD galaxies, a particular type of stars appears: they are characterized by [C/Fe] $\geqslant$ +0.7 dex \citep{aoki2007}, therefore they are called Carbon-Enhanced Metal-Poor (CEMP) stars.
Since their origin has not been clearly understood and their abundances are very peculiar they are not considered in the following analysis.

\begin{table*}
%\centering
\hspace*{-0.5cm}
\caption{Physical features of UFDs. \textit{Columns:} (1) name of the galaxy; (2) absolute magnitude in V-band; (3) half-light radius; (4) distance;
(5) surface brightness; (6) mean [Fe/H] value; (7) dispersion in [Fe/H]; (8) mass to light ratio; (9) and (10) present-day stellar mass derived assuming the Salpeter and the \citet{kroupa1993} IMF, respectively. \textit{References:} columns (2), (4), (6) and (7) \citet{simon2019}; (3), (9) and (10) \citet{martin2008}; (5) \citet{munoz2018}; (8)  \citet{collins2014} for Boo I and CVn I and \citet{walsh2008} for Boo II.}
\begin{tabular}{cccccccccc}
\hline
\hline
UFD&$M_V$&$R_{1/2}$&$D$&$\mu_V$&$\langle\mathrm{[Fe/H]}\rangle$&$\sigma_{\mathrm{[Fe/H]}}$&$(M/L)_V$&$M^{Salpeter}_{\star}$&$M^{Kroupa}_{\star}$\\ 
&(mag)&(pc)&(kpc)&$(\mathrm{mag \cdot arcsec^{-2}})$&(dex)&(dex)&$(\mathrm{M_{\odot}/L_{\odot}})$&$(\mathrm{M_{\odot}})$&$(\mathrm{M_{\odot}})$\\[0.07cm]\hline
\rule{0pt}{1.\normalbaselineskip}
\hspace{-0.15cm}
Boo I &$-6.02\pm0.25$&242$^{+22}_{-20}$&$66.0\pm 3.0$&$28.4\pm0.31$&$-2.35^{+0.09}_{-0.08}$&0.44$^{+0.07}_{-0.06}$&$198.0^{+83.4}_{-69.1}$&$(6.7\pm0.6)\cdot 10^4$&$(3.4\pm0.3)\cdot 10^4$\\[0.2cm] 
Boo II &$-2.94^{+0.74}_{-0.75}$&$51\pm17$&$42.0\pm1.0$&27.56$^{+1.04}_{-1.08}$&$-2.79^{+0.06}_{-0.10}$&<0.35&$98 ^{+420}_{-84}$&$(2.8^{+0.7}_{-0.5})\cdot 10^3$&$(1.4^{+1.3}_{-1.0})\cdot 10^3$\\[0.2cm] 
CVn I&$-8.73\pm0.06$&$564\pm36$&$211.0\pm6.0$&$27.1\pm0.19$&$-1.91\pm0.04$&0.39$^{+0.03}_{-0.02}$&$164.3\pm31.2$&$(5.8\pm0.4)\cdot 10^5$&$(3.0\pm0.2)\cdot 10^5$\\
\hline
\hline
\end{tabular}

\label{tab:ufdsfetaures}
\end{table*}

\subsection{Bo\"otes I}
Bo\"{o}tes I was discovered from the analysis of the SDSS DR5 images by \cite{belokurov2006}. Through CMD fitting analysis \cite{okamoto2012} derived that the distribution of stars are well fitted by an isochrone of age around 13.7 Gyr and [Fe/H] near $-2.3$ dex. 

The chemical abundance data have been taken from the works of \cite{Feltzing}, \cite{norris2010high}, \citet['GM' analysis]{gilmore} and \cite{ishigaki2014}. For the MDF we have added the non-overlapping stars of \cite{martin2007}, \cite{norris2010low} and \cite{lai2011}. 
%Concerning instead the observed stellar mass, the third constraint we have, \citet{martin2008} derived $M_{\star}^{Salpeter}=\ (6.7\pm0.6)\ \cdot 10^4\ \mathrm{M_{\odot}}$ and $M_{\star}^{Kroupa}=\ (3.4\pm0.3)\ \cdot 10^4\ \mathrm{M_{\odot}}$ \footnote{The Kroupa IMF adopted in \cite{martin2008} is the one proposed by \cite{kroupa1993}} for Bo\"otes I. 

\subsection{Bo\"otes II}
Bo\"otes II was discovered by \cite{walsh2007} as a resolved stellar overdensity in an automated search of the SDSS DR5 imaging data.

% Using a bootstrap approach, \cite{walsh2008} derived physical parameters of Bo\"otes II from the data collected through the MegaCam mounted on the Multiple Mirror Telescope (MMT). They derived an absolute magnitude of $M_{V,tot}$ = $-2.4\ \pm$ 0.7 mag and a half-light radius of 36 $\pm$ 9 kpc using a Plummer profile. Comparing the CMD of Bo\"otes II with the ones of several globular clusters they estimated the distance modulus corresponding to a distance of D = 42 $\pm$ 2 kpc. To derive the age, instead, they used the isochrone fitting: the best fit has been obtained with [Fe/H]= $-2.3$ dex and an age of 13 Gyr.

The data samples of chemical abundances have been taken from \cite{koch2014} and \cite{ji2016} while for the construction of the MDF we added the data of \cite{koch2009}.
%The present-day stellar mass of Bo\"otes II has been estimated by \cite{martin2008} to be $M_{\star}^{Salpeter}=\ 2.8^{+0.7}_{-0.5}\ \cdot 10^3\ \mathrm{M_{\odot}}$ and $M_{\star}^{Kroupa}=\ 1.4^{+1.3}_{-1.0}\ \cdot 10^3\ \mathrm{M_{\odot}}$.

\subsection{Canes Venatici I}

\cite{zucker2006} discovered the UFD galaxy Canes Venatici I in the SDSS DR5 data.

%They derived an absolute magnitude of $M_{V,tot}$ = $-7.9$ mag. \cite{okamoto2012} derived the distance of Canes Venatici I measuring the apparent magnitude of the HB and the Tip of the Red Giant Branch (TRGB) finding a good agreement with the result obtained by \cite{kuehn2008} who estimated the magnitude of 23 RR Lyrae stars. The derived distance is D = 216 $\pm$ 8 kpc. Moreover, studying the CMD, \cite{okamoto2012} found that the system has an average metallicity of $\langle\mathrm{[Fe/H]}\rangle$ = $-2.3$ dex and an age of 12.6 Gyr. They estimated also a half-light radius of $\sim$ 590 pc if a Plummer profile is assumed.

Recently, \citet{munoz2018} carried out an imaging survey of the outer halo satellites deriving an absolute magnitude of $M_{V,tot}$=-8.8 mag which places CVn I as a dSph galaxy. Moreover, \citet{weisz2014}, analyzing the data collected with the Hubble Space Telescope (HST), derived that the star formation in CVn I lasted 5 Gyr, longer than the typical SFH of UFDs.

The chemical abundance data we used are taken from \cite{francois2016}, who analyzed only two stars in this galaxy, while, for the MDF, we have added the samples of \cite{martin2007} and \cite{kirby2010}; this last work provides the [Fe/H] value for 174 member stars for CVn I.
%The observed stellar mass of Canes Venatici I, derived by \cite{martin2008}, is equal to $M_{\star}^{Salpeter}=\ (5.8\pm0.4)\ \cdot 10^5\ \mathrm{M_{\odot}}$ and $M_{\star}^{Kroupa}=\ (3.0\pm0.2)\ \cdot 10^5\ \mathrm{M_{\odot}}$.

\medskip
In Table \ref{tab:ufdsfetaures} are summarized the updated observational features of the three UFDs we have analyzed in this work. The stellar masses have been derived by \citet{martin2008} assuming two different IMFs: Salpeter and Kroupa IMFs \citep{Salpeter1955,kroupa1993}. We report these values in Table \ref{tab:ufdsfetaures} where one can see that the differences between these masses are small and inside a factor of two. Unfortunately, the IGIMF has never been used to derive the stellar mass of these galaxies; therefore, at the beginning our analysis, all the models that predict a present-day stellar mass between these two values (errors included) have been considered as good models. Then, the outputs of such models have been coupled with the PARSEC stellar isochrones \citep{bressan2012,tang2014,chen2015} in our photochemical model \citep{vincenzo2016} in order to predict the visual magnitudes. We have finally compared the predicted magnitudes with the observed ones (taken from the same study of the present-day stellar mass) to verify whether the good models previously selected were still able to fit the new, and IMF-independent, observational constraint.

%\subsection{Hercules}
%
%Hercules was discovered by \cite{belokurov2007} from the SDSS DR5 imaging data. They estimated an absolute magnitude of $M_{V,tot}$ = $-6.0\ \pm$ 0.6 mag, a distance of $140 \pm 13$ pc and a half-light radius of $\sim$ 320 pc for a Plummer profile. \cite{sand2009}, using the color-magnitude-fitting package StarFISH, have determine that the stellar populations in Hercules are older than 12 Gyr and metal-poor ([Fe/H] $\sim$ $-2.0$ dex). They were able also to search for external structures and what they found are evidences that Hercules is embedded in a larger stream of stars.
%
%The data set of chemical abundances we have used to compare with the model predictions has been built collecting the data from \cite{koch2008}, \cite{aden2011} and \cite{francois2016} while for the MDF we have added the data from \cite{aden2009}and \cite{vargas2013}.

\section{Results}
\label{sec:results}
 
For every galaxy, the mass of the dark matter halo $M_{DM}$, the half-light radius $r_L$ and the star formation history (SFH) we have adopted are derived observationally and are maintained fixed for all the models we have ran. On the other hand, the star formation efficiency and the infall mass are varied in order to reproduce the observational constraints. Our method is to impose a final stellar mass for each galaxy and consider the gas abundances as the unknowns of the problem.

\subsection{Chemical evolution of Bo\"otes I}
We have assumed a dark matter halo of $M_{DM}=3.0\cdot 10^6\ \mathrm{M_{\odot}}$ \citep{collins2014}, while for the effective radius of the luminous (baryonic) component we have adopted the value estimated by \cite{martin2008} of $r_L=242$ pc. The star formation history of Boo I has been derived from the CMD fitting analysis by \cite{brown2014}. They estimate that the stars have been formed in 1 Gyr, a quite different result from the previous estimatation done by \cite{dejong2008} of 4 Gyr. However, \cite{brown2014} used more precise data coming from the HST and for this reason we have preferred their estimation.

\begin{table*}
\centering
\hspace*{-0.5 cm}
\caption{Input parameters used for all the chemical evolution models performed for Bo\"otes I. \textit{Columns:} (1) star formation efficiency, (2) wind efficiency, (3) infall time-scale, (4) star formation history \citep{brown2014}, (5) total infall gas mass, (6) mass of the dark matter halo \citep{collins2014} obtained using \citet{martin2008} half-light radius values, (7) half-light radius \citep{martin2008}, (8) ratio between the half-light radius and the dark matter effective radius, (9) initial mass function.}

\begin{tabular}{ccccccccc}
\hline
\hline
\multicolumn{9}{c}{ \bf Bo\"otes I: parameters of the models} \\
\rule{0pt}{1.\normalbaselineskip}
$\nu$&$\omega$&$\tau_{inf}$&SFH&$M_{infall}$&$M_{DM}$&$r_L$&$S=\frac{r_L}{r_{DM}}$&IMF \\
$\mathrm{(Gyr^{-1})}$&$\mathrm{(Gyr^{-1})}$&$\mathrm{(Gyr)}$&$\mathrm{(Gyr)}$&$\mathrm{( M_{\odot})}$&$\mathrm{( M_{\odot})}$&$\mathrm{(pc)}$&\\[0.1cm] \hline
\noalign{\vskip 0.065in} 
0.005/0.01/0.1&10&0.005&$0-1$&$1.0/2.5\cdot 10^7$&$3.0\cdot 10^6$&$242$&0.3&IGIMF/Salpeter\\
\hline
\hline
\end{tabular}

\label{tab:databoo1}
\end{table*}

In Table \ref{tab:databoo1} are reported the input parameters of the most relevant models we obtained for the chemical evolution of Boo I. We tested models with three different star formation efficiencies: 0.005, 0.01 and 0.1 $\mathrm{Gyr^{-1}}$. The first two, as derived also by \cite{vincenzo2014}, are the most likely values for these types of galaxies. Only assuming such low star formation efficiencies we are able to explain the observed decline of [$\alpha$/Fe] abundance ratios at very low [Fe/H] for a Salpeter IMF. Lowering the $\nu$ parameter, in fact, leads to a lower production of iron from CC SNe. Consequently, the [Fe/H] value at which Type Ia SNe start polluting the gas in the ISM decreases. The 0.1 $\mathrm{Gyr^{-1}}$ SFE, instead, is more typical of dSph galaxies which are supposed to have experienced a longer and more intense star formation; moreover, the [Fe/H] value at which the decline of the [$\alpha$/Fe] ratios starts is higher than in the case of UFDs. This behaviour reflects the time-delay model for the chemical enrichment in different regimes of SFR \citep{matteucci2012}. Nevertheless, we have also tested $\nu=0.1\mathrm{Gyr^{-1}}$ in order to reproduce better the [$\alpha$/Fe] trends and the MDF, as we will explain later.

Concerning the infall mass $M_{infall}$, we have considered two values: $1.0\cdot 10^7\  \mathrm{M_{\odot}}$ and $2.5\cdot 10^7\  \mathrm{M_{\odot}}$ in order to reproduce the observed present-day stellar mass of Boo I. The chosen infalling mass is always larger than the final stellar mass since part of the gas is lost through galactic winds. The infalling gas has been assumed to be primordial and that it has been accreted by the potential well of the dark matter halo in a very short time; consequently the infall time-scale has been set to $\tau_{inf}$ = 0.005 Gyr.  

Finally, for every value of SFE and infall mass we have ran two models: one adopts the Salpeter IMF while the other the IGIMF proposed by \citetalias{recchi}  in order to compare the results.

\begin{figure} 
 %\centering
 \hspace{-0.5cm}
  \includegraphics[width=1.05\columnwidth]{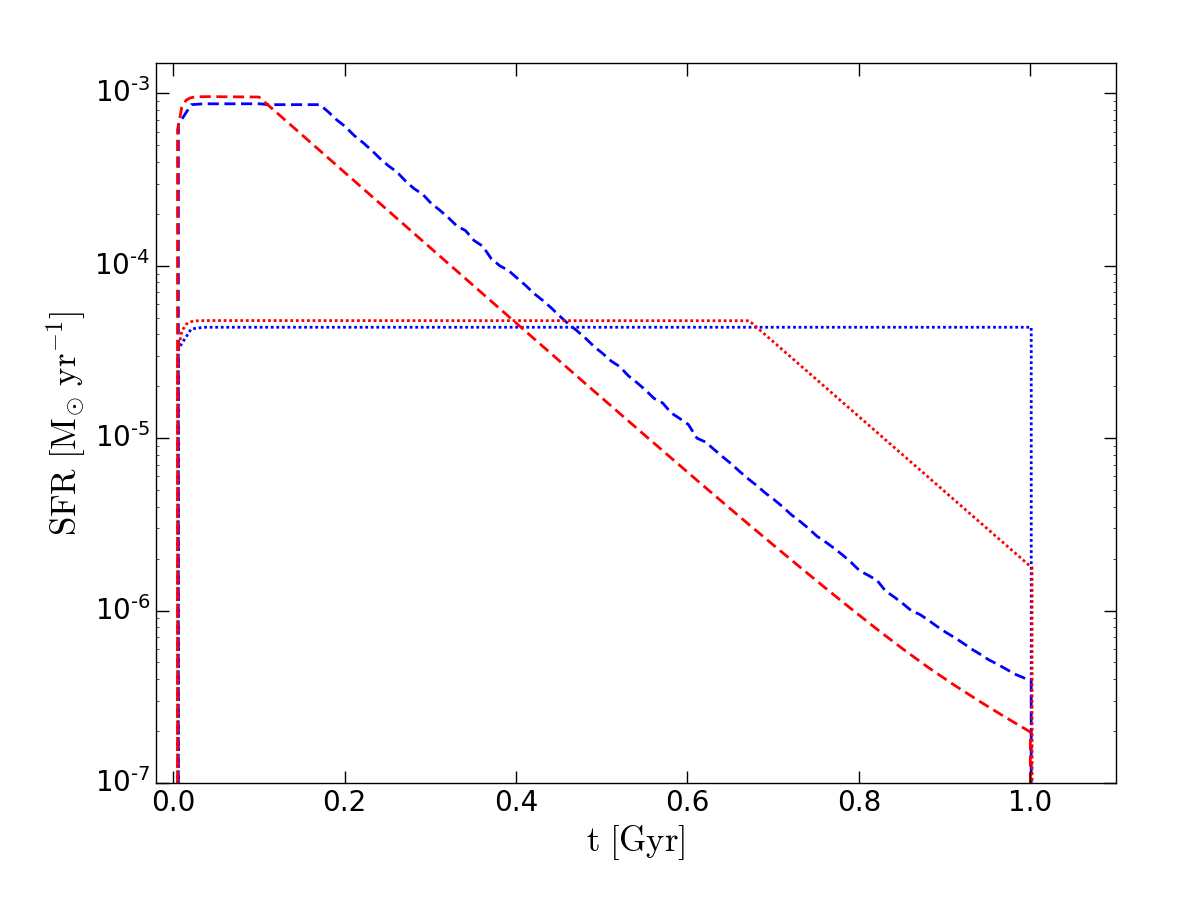}
  \caption{The star formation rate is shown as a function of time for $M_{infall}=1.0 \cdot 10^7\  \mathrm{M_{\odot}}$. The red lines represent the models with the Salpeter IMF while the blue lines refer to the results obtained with the IGIMF. With the dotted lines are shown the models for $\nu=0.005\ \mathrm{Gyr^{-1}}$ while with the dashed lines the ones for $\nu=0.1\ \mathrm{Gyr^{-1}}$. }
  \label{fig:sfrboo1}
\end{figure}

In Figure \ref{fig:sfrboo1} are shown the predictions of the SFR of Boo I. The models are obtained with both the IMF parametrizations: the results for the Salpeter IMF are presented in red, while for the IGIMF in blue. The infall mass has been set to $M_{infall}=1.0 \cdot 10^7\  \mathrm{M_{\odot}}$ and, for the SFE, we have chosen to plot the two extreme values: $\nu=0.005\ \mathrm{Gyr^{-1}}$ and $\nu=0.1\ \mathrm{Gyr^{-1}}$.

One can see in Figure \ref{fig:sfrboo1} the declines of the SFR before 1 Gyr: they are caused by the decrease of the amount of gas in the ISM after the onset of the galactic wind (the time of its appearance is reported in Table \ref{tab:boo1res}). Moreover, we can also see that the SFRs are flat before the onset of the wind; actually, the SFR is not constant but it slightly decreases, given the very small mass of gas consumed to form stars. Comparing the models with different IMFs, it emerges that the Salpeter IMF predicts a higher SFR than the IGIMF. This difference can be explained by the higher number of low mass stars predicted assuming the IGIMF. In fact, low mass stars lock up gas which cannot be used to form new stars.

The models obtained with $M_{infall}=\ 2.5 \cdot 10^7\ \mathrm{M_{\odot}}$ predict higher SFR given the proportionality between the SFR and the mass of the gas. As reported in Table \ref{tab:boo1res} the onset of the galactic wind in these models starts later than for $M_{infall}=\ 1.0 \cdot 10^7\  \mathrm{M_{\odot}}$, in some cases even after the end of the star formation.

\begin{figure} 
 %\centering
 \hspace{-0.5cm}
  \includegraphics[width=1.\columnwidth]{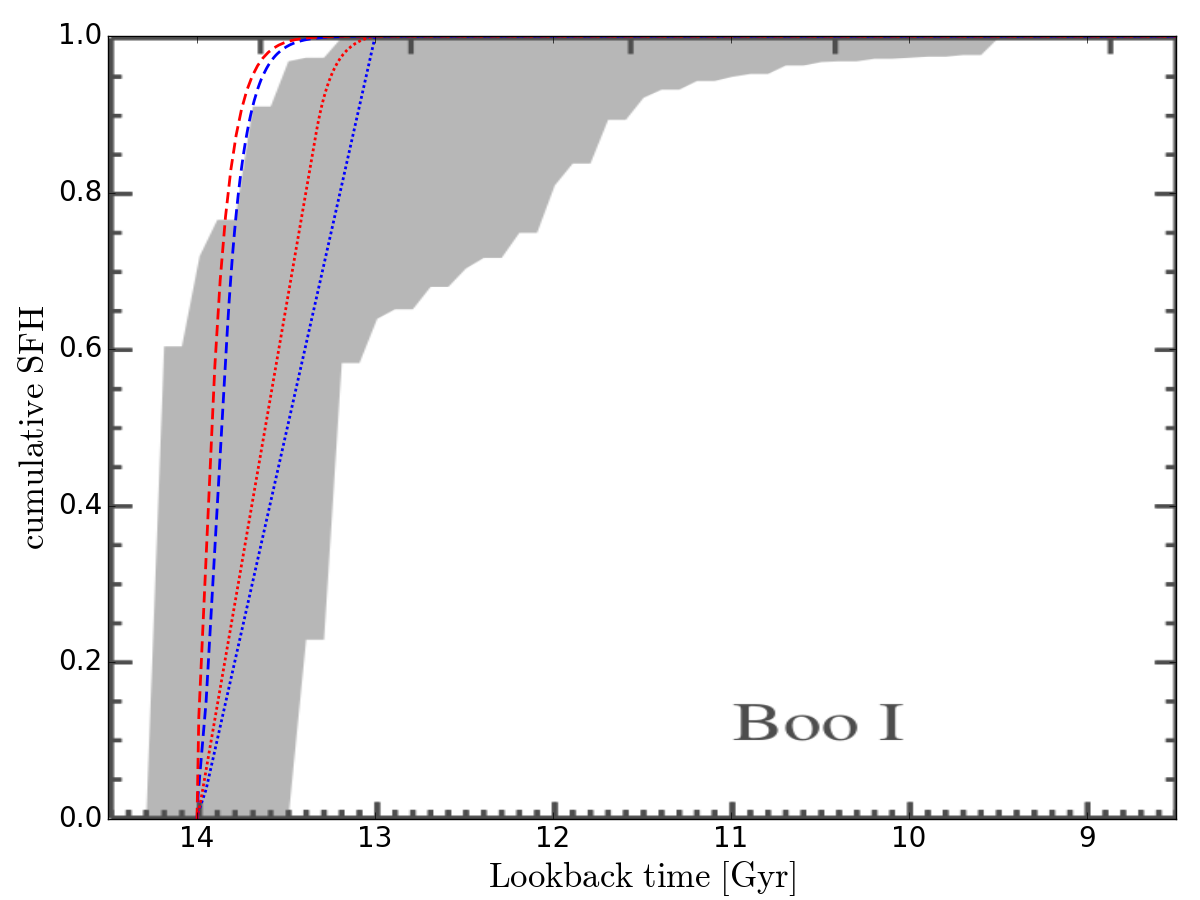}
  \caption{The cumulative star formation rate is shown as a function of time for $M_{infall}=1.0 \cdot 10^7\  \mathrm{M_{\odot}}$. The red lines represent the models with the Salpeter IMF while the blue lines refer to the results obtained with the IGIMF. With the dotted lines are shown the models for $\nu=0.005\ \mathrm{Gyr^{-1}}$ while with the dashed lines the ones for $\nu=0.1\ \mathrm{Gyr^{-1}}$. In grey the statistical uncertainties for the cumulative SFH taken from \citet{brown2014} are shown.}
  \label{fig:cumsfrboo1}
\end{figure}

In Figure \ref{fig:cumsfrboo1} are shown the predicted cumulative SFRs compared with the statistical uncertainties for the cumulative SFH obtained from photometric and spectroscopic data by \citet{brown2014}. The models with low SFE are well within the uncertainties confirming that the star formation in these systems is quite inefficient.  

\begin{figure} 
 %\centering
 \hspace{-0.5cm}
  \includegraphics[width=1.05\columnwidth]{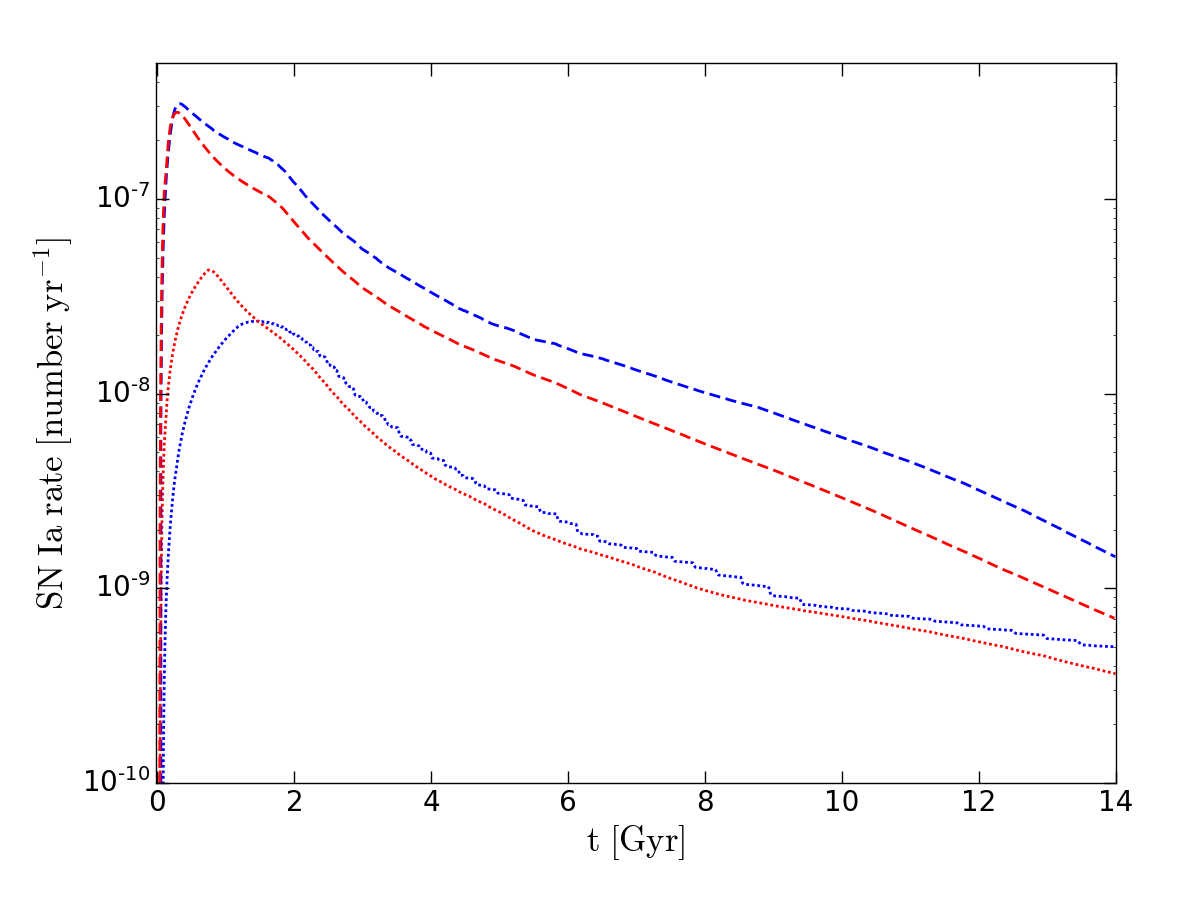}
  \caption{The Type Ia supernova rate is shown as a function of time for $M_{infall}=1.0 \cdot 10^7\  \mathrm{M_{\odot}}$. The red lines represent the models with the Salpeter IMF while the blue lines refer to the results obtained with the IGIMF. With the dotted lines are shown the models for $\nu=0.005\ \mathrm{Gyr^{-1}}$ while with the dashed lines the ones for $\nu=0.1\ \mathrm{Gyr^{-1}}$.}
  \label{fig:sne1arateboo1}
\end{figure}

In Figure \ref{fig:sne1arateboo1} are shown the rates of Type Ia SNe for the same models of Figure \ref{fig:sfrboo1}. For high SFE the IGIMF predicts a higher number of Type Ia supernova progenitors than the Salpeter one. 
On the other hand, at early times and for low SFE, the IGIMF predicts a lower number of Type Ia SNe, as shown for $\nu=0.005\ \mathrm{Gyr^{-1}}$. Consequently, a lower number of binary systems leading to SNe Ia in the highest mass range are formed compared to what obtained with the Salpeter IMF. 

%The effect of the strong truncation of the IGIMF can be seen more clearly in the models with $\nu=0.005\ \mathrm{Gyr^{-1}}$; in this case, the SFR is smaller than $10^{-4}\ \mathrm{M_{\odot}  yr^{-1}}$ (see Figure \ref{fig:sfrboo1}) and for these values the IGIMF predicts a truncation at masses lower than $10\mathrm{M_{\odot}}$, as it can be seen in Figure \ref{igimf2}. For this reason, at the very beginning, only assuming the Salpeter IMF it is possible to form binary systems with high masses which are responsible for the peak of the SN rate at very early times.

%At early times it is not true since the most massive binary systems ($M_{binary}=M_1+M_2 \sim 16 \mathrm{M_{\odot}}$) explode, but, even for $\nu=0.1\ \mathrm{Gyr^{-1}}$, the SFR is very low implying a truncation of the IGIMF at lower masses. This heavy truncation leads to a lower number of binary systems with the highest mass than the ones predicted with the Salpeter IMF and consequently to a lower number of Type Ia SN explosions. The effect of the strong truncation of the IGIMF can be seen more clearly in the models with $\nu=0.005\ \mathrm{Gyr^{-1}}$; in this case, the SFR is smaller than $10^{-4}\ \mathrm{M_{\odot}  yr^{-1}}$ (see Figure \ref{fig:sfrboo1}) and for these values the IGIMF predicts a truncation at masses lower than $10\mathrm{M_{\odot}}$, as it can be seen in Figure \ref{igimf2}. For this reason, at the very beginning, only assuming the Salpeter IMF it is possible to form binary systems with high masses which are responsible for the peak of the SN rate at very early times.

\begin{figure} 
 %\centering
 \hspace{-0.5cm}
  \includegraphics[width=1.05\columnwidth]{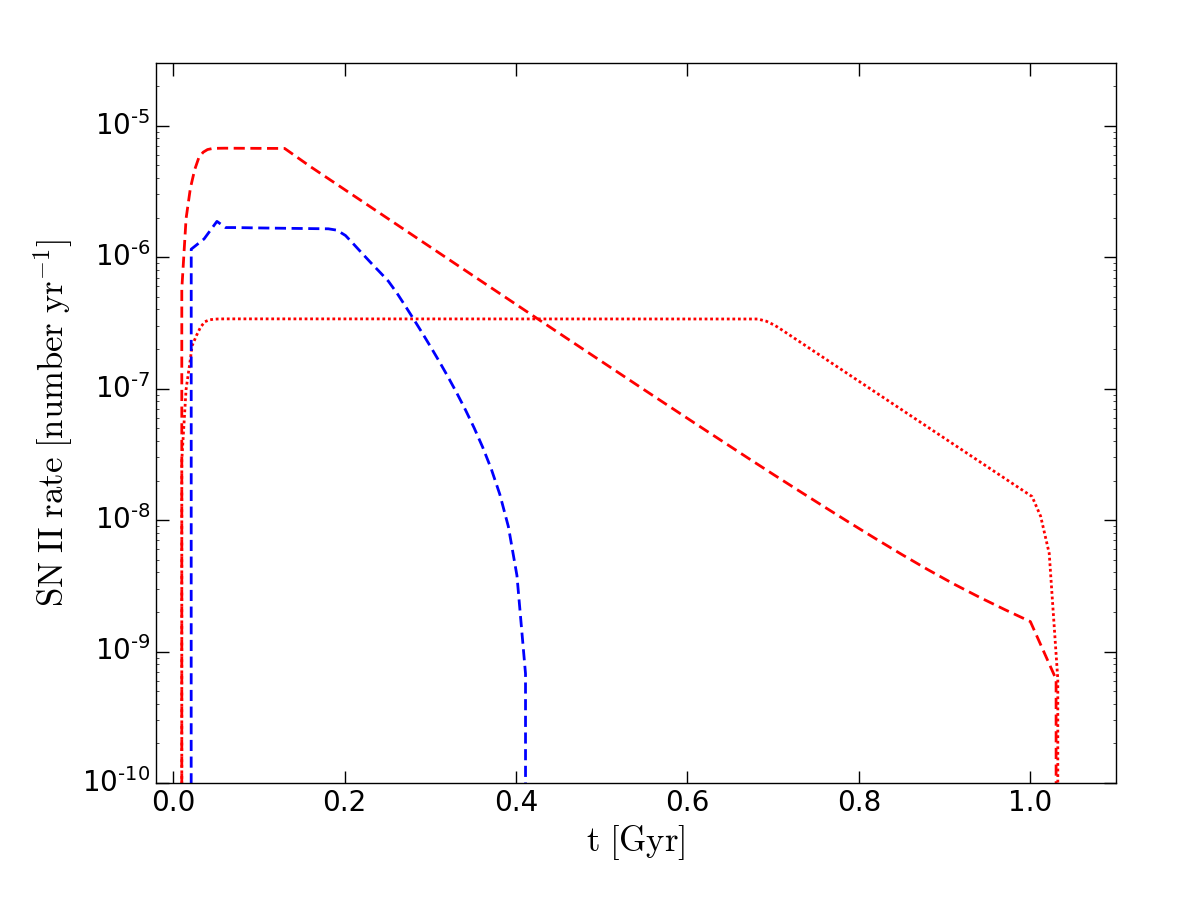}
  \caption{The Type II supernova rate is shown as a function of time for $M_{infall}=1.0 \cdot 10^7\  \mathrm{M_{\odot}}$. The red lines represent the models with the Salpeter IMF while the blue lines refer to the results obtained with the IGIMF. The models for $\nu=0.005\ \mathrm{Gyr^{-1}}$ are shown with the dotted lines, while with the dashed lines the ones for $\nu=0.1\ \mathrm{Gyr^{-1}}$. }
  \label{fig:sn2rateboo1}
\end{figure}

Figure \ref{fig:sn2rateboo1} shows the Type II supernova rate predicted for Boo I. The models  are the same as Figure \ref{fig:sfrboo1} and \ref{fig:sne1arateboo1}. 
The trends shown here are similar to the ones obtained for the SFR, especially for the models adopting a Salpeter IMF. This similarity can be explained by the short lifetimes of the progenitors of Type II SNe. Thus, the rate of their explosions follows the rate at which the stars are formed. However, the same cannot be said for the models with the IGIMF. The model with $\nu=0.005\ \mathrm{Gyr^{-1}}$ do not even predict the existence of Type II SNe because of the very low SFR. On the contrary, the model with $\nu=0.1\ \mathrm{Gyr^{-1}}$ follows the trend of the SFR until the galactic wind reduces the SFR at a level at which the truncation of the IGIMF is so strong that no stars with M $\geq 6-8 \mathrm{M_{\odot}}$ are formed.

\begin{table*}

\centering
%\hspace*{-2cm}
\caption{ The predictions of chemical evolution models for Bo\"otes I. The second and the third  column contain the input parameters which have been varied in the models while the other six concern the model predictions. \textit{Columns:} (1) model name, (2) infall mass, (3) star formation efficiency, (4) present-day stellar mass derived with the IGIMF, (5) visual magnitude derived with the IGIMF, (6) present-day stellar mass derived with the Salpeter IMF, (7) time of the onset of the galactic wind assuming the IGIMF, (8) time of the onset of the galactic wind assuming the Salpeter IMF, (9) peak of the stellar MDF obtained with the IGIMF, (10) peak of the stellar MDF obtained with the Salpeter IMF.}

 \resizebox{1.85\columnwidth}{!}{
 \begin{tabular}{ccc|ccccccc}
\hline
\hline

\multicolumn{10}{c}{{ \bf Bo\"otes I: model predictions}} \\[0.2cm]
\rule{0pt}{1.3\normalbaselineskip}

%Input parameters&&&Model Predictions\\
Model name&$M_{infall}$&$\nu$&$M_{\star,fin}^{IGIMF}$&$M_V^{IGIMF}$&$M_{\star,fin}^{Salpeter}$&$t_{wind}^{IGIMF}$&$t_{wind}^{Salpeter}$&$\mathrm{[Fe/H]_{peak}^{IGIMF}}$&$\mathrm{[Fe/H]_{peak}^{Salpeter}}$\\[0.2cm]
&($\mathrm{M_{\odot}} $)&($\mathrm{Gyr^{-1}} $)&($\mathrm{M_{\odot}} $)&(mag)&($\mathrm{M_{\odot}} $)&(Gyr)&(Gyr)&(dex)&(dex)\\[0.17cm] \hline   
\rule{0pt}{1.\normalbaselineskip}
\hspace{-0.15cm}
1BooI&$1.0 \cdot 10^{7}$&0.005&$3.5 \cdot 10^{4}$&$-5.89$&$2.2 \cdot 10^{4}$&1.54&0.68&-3.3&-2.7\\  
2BooI&$1.0 \cdot 10^{7}$&0.01&$6.6 \cdot 10^{4}$&$-6.51$&$3.1 \cdot 10^{4}$&0.87& 0.49&-2.7&-2.7\\
3BooI&$1.0 \cdot 10^{7}$&0.1&$1.7 \cdot 10^{5}$&&$1.1 \cdot 10^{5}$&0.18&0.13&-2.7&-2.3\\ \
&&&&&&&\\
4BooI&$2.5 \cdot 10^{7}$&0.005&$8.9 \cdot 10^{4}$&&$6.8 \cdot 10^{4}$&2.07&1.33&-2.9&-2.5\\  
5BooI&$2.5 \cdot 10^{7}$&0.01&$1.8 \cdot 10^{5}$&&$1.2 \cdot 10^{5}$&1.19&0.82&-2.5&-2.3\\
6BooI&$2.5 \cdot 10^{7}$&0.1&$6.3 \cdot 10^{5}$&&$3.9 \cdot 10^{5}$&0.27&0.22&-2.1&-2.1\\
\hline
\hline
\end{tabular}
}

\label{tab:boo1res}
\end{table*}

In Table \ref{tab:boo1res} we present the predictions of our chemical evolution models. In the second and third column we have listed the input parameters we have varied: $M_{infall}$ and $\nu$. Low values of the star formation efficiencies lead to a lower [Fe/H] at which the MDF reaches its peak, since the stars pollute more slowly the ISM, arriving at the quench of the SF at low [Fe/H] values. Also the present-day stellar mass is influenced by the SFE in a similar way: the slower the production of stars the lower the mass in stars when the SF stops, leading to a lower stellar mass at the present time. A low SFE causes also a later onset of the galactic wind, given the related decrease of the number of SN events which heat up the ISM. 
When, instead, we vary the infall mass, the immediate consequence is the variation of the present-day stellar mass. Moreover, an increase of the infall mass makes the gas more bound to the galaxy, implying a later onset of the galactic wind. Therefore, the SFR starts decreasing later, allowing the formation of a higher number of stars at high [Fe/H] values which shifts the peak of the MDF towards higher metallicities.

Comparing the observed values $M_{\star}^{Salpeter}=\ (6.7\pm0.6)\ \cdot 10^4\ \mathrm{M_{\odot}}$ and $M_{\star}^{Kroupa}=\ (3.4\pm0.3)\ \cdot 10^4\ \mathrm{M_{\odot}}$ obtained by \cite{martin2008} with the predicted ones, we see that, assuming the IGIMF, the models 1BooI and 2BooI are in agreement with the observed values. However, the observed masses were not derived assuming the IGIMF, therefore we have applied our photochemical model \citep{vincenzo2016} to derive, for 1Boo1 and 2Boo1, the predicted visual magnitudes. For 1BooI, the accordance with the observed value of $-6.02\pm0.25 $ mag is within the observational error, at variance with the 2BooI model. Therefore, we consider 1BooI the best model in order to fit the stellar mass. Concerning the Salpeter IMF, instead, the present day stellar mass can be reproduced by the models 2BooI and 4BooI.

The results obtained with the two different IMF parametrizations suggest that, generally, the IGIMF predicts higher present-day stellar masses, a later onset of the galactic wind and more metal-poor stars than the Salpeter IMF.
The later onset of the galactic wind can be explained by the lower production of massive stars by the IGIMF leading to a lower number of core-collapse SNe explosions. On the contrary, the IGIMF predicts a higher amount of gas blocked in very low mass stars which is released on long time-scales causing a higher present-day stellar mass. High amounts of low mass stars also induce a lower reprocessment of the gas, thus the MDF is peaked at lower [Fe/H] values.

%Even though enhancing the SFE the IGIMF becomes more similar to the Salpeter IMF, in the case of $\nu=0.1\ \mathrm{Gyr^{-1}}$ and $M_{infall}=1.0\cdot 10^7\mathrm{M_{\odot}}$, the [Fe/H]$^{IGIMF}_{peak}$ do not follow the increase derived for the Salpeter IMF. In addiction, its value decreases with respect to the model for $\nu=0.01\ \mathrm{Gyr^{-1}}$. The explanation lies in the very early onset of the galactic wind which diminuish the SFR: consequently the majority of the stars form at very low [Fe/H]. This effect is also summed to the slowering of the iron production since, for the IGIMF, a decrease of the SFR induces a lower production of massive stars which pollute the ISM in very short time-scales.

\subsubsection{Bo\"otes I: Abundance ratios and its interpretation}
From the dataset of [$\alpha$/Fe] abundance ratios  we have selected, we can argue that most of the stars are very metal-poor and are enhanced in $\alpha$-elements. Moreover, the higher their [Fe/H] value the lower is their [$\alpha$/Fe] ratio. This is basically true for all the stars except for a star in \cite{gilmore} sample. This star has a [Fe/H] = $-1.80$ dex and displays high abundances, in particular of Ti and Mg.

\begin{figure*} 
 %\centering
 %\hspace{-2.15cm}
  \includegraphics[width=0.85\textwidth]{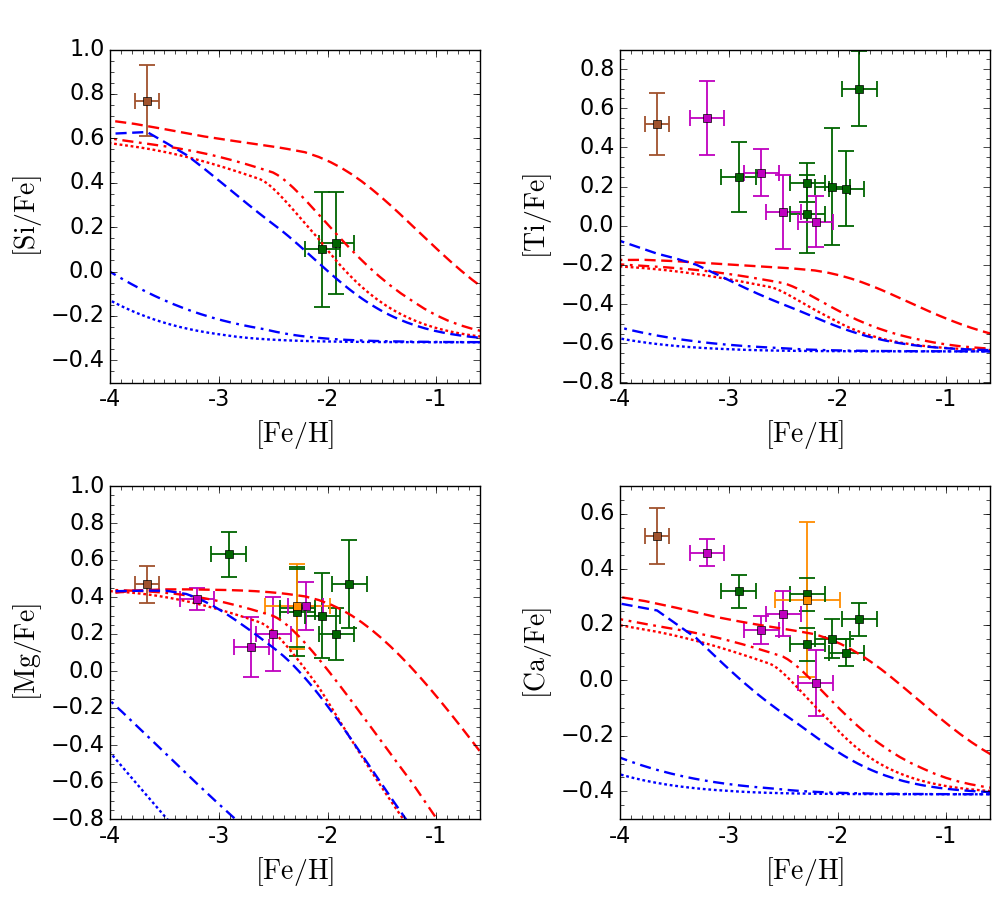}
  \caption{In this figure we compare the $[\alpha$/Fe] versus [Fe/H] abundance ratios for Si, Mg, Ti and Ca as observed in Bootes I UFD member stars with the predictions of the chemical models with $\omega =\ 10 \ \mathrm{Gyr^{-1}}$, $M_{infall}=1.0 \cdot 10^7\  \mathrm{M_{\odot}}$ for different values for $\nu$ and varying the IMF. In red we show the models with the Salpeter IMF while in blue the ones with the IGIMF. The models with $\nu=0.005\ \mathrm{Gyr^{-1}}$ (1BooI) are represented by the dotted line, the ones with $\nu=0.01\ \mathrm{Gyr^{-1}}$ (2BooI) by the dash-dotted line, while the models with $\nu=0.1\ \mathrm{Gyr^{-1}}$ (3BooI) are plotted with the dashed line. Concerning the data, the green squares refer to the data sample of \citet{gilmore}, the magenta ones of \citet{Ishigaki}, the orange one of \citet{Feltzing}, while the brown one is taken from \citet{norris2010high}.}
  \label{fig:abundboo1}
\end{figure*}

In Figure \ref{fig:abundboo1} we show the effects of changing the SFE and the IMF parametrization on the [$\alpha$/Fe] vs. [Fe/H] trends for the models with $M_{infall}=\ 1.0\cdot 10^7\ \mathrm{M_{\odot}}$ which best reproduce the present-day stellar mass. The models with the Salpeter IMF (red lines) predict a decrease of the [$\alpha$/Fe] at higher [Fe/H] than the models with the IGIMF (blue lines). This is due to the strong suppression of massive stars at low SFRs ($10^{-4}-10^{-5}\ \mathrm{M_{\odot}yr^{-1}}$) when adopting the IGIMF. Therefore, the results obtained using the Salpeter IMF, in particular the models 1BooI and 2BooI, are able to reproduce better the observed abundances with respect to the ones with the IGIMF. The only exception is the Ti abundance which cannot be reproduced by any of the models; this discrepancy can be explained given the large uncertainties of the available yields for this element (see \citealp{romano2010}). Concerning the IGIMF, the model which best reproduces the observational data is 3BooI, the one with $\nu=0.1\ \mathrm{Gyr^{-1}}$ (dashed blue line), even though it underestimates the [Ca/Fe] ratio. The other two models with the IGIMF, instead, predict the decrease of [$\alpha$/Fe] at extremely low metallicities ([Fe/H] $\sim -5$ dex) not in agreement with data.

\begin{figure*} 
 %\centering
 %\hspace{-2.15cm}
  \includegraphics[width=0.85\textwidth]{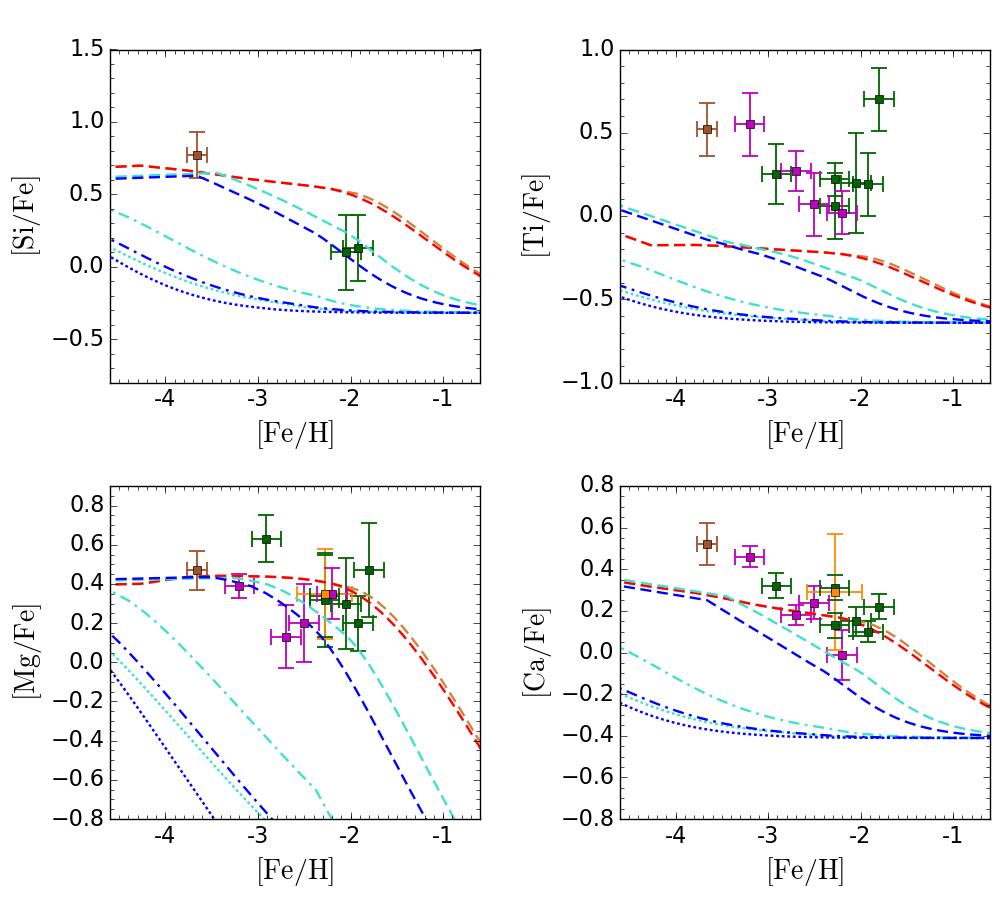}
  \caption{In the figure we compare the $[\alpha$/Fe] versus [Fe/H] abundance ratios for Si, Mg, Ti and Ca as observed in Bootes I UFD member stars with the predictions of the chemical models for different values of $\nu$ and $M_{infall}$, varying also the IMF. In red and brown we show the models 3BooI and 6BooI, respectively, assuming a Salpeter IMF. The blue and cyan lines represent the models adopting the IGIMF for $M_{infall}=1.0 \cdot 10^7\ \mathrm{M_{\odot}}$ and $M_{infall}=2.5 \cdot 10^7\ \mathrm{M_{\odot}}$, respectively. The dotted lines refer to $\nu=0.005\ \mathrm{Gyr^{-1}}$, the dash-dotted lines to $\nu=0.01\ \mathrm{Gyr^{-1}}$ and the dash-dotted lines to $\nu=0.1\ \mathrm{Gyr^{-1}}$.}
  \label{fig:abundboo1mass}
\end{figure*}

In Figure \ref{fig:abundboo1mass} we show the results obtained increasing the infall mass to $M_{infall}=2.5 \cdot 10^7\  \mathrm{M_{\odot}}$ (brown lines for the Salpeter IMF, cyan for the IGIMF). The model adopting the Salpeter IMF does not suffer a substantial variation as the ones with the IGIMF; this is caused by the dependency of the IGIMF on the SFR. A higher infall mass increases the SFR which induces a shift of the truncation of the IGIMF towards higher masses. This means that more CC SNe are expected to occur before the appearance of Type Ia SNe, shifting the knee of the [$\alpha$/Fe] at higher [Fe/H]. Nevertheless, the increase of the SFR is not enough to allow the models with $\nu=0.005\ \mathrm{Gyr^{-1}}$ and $\nu=0.01\ \mathrm{Gyr^{-1}}$ to fit the data even for this infall mass.

\subsubsection{Bo\"otes I: MDF and its interpretation}
\begin{figure*} 
 %\centering
% \hspace{-0.7cm}
  \includegraphics[width=1.\textwidth]{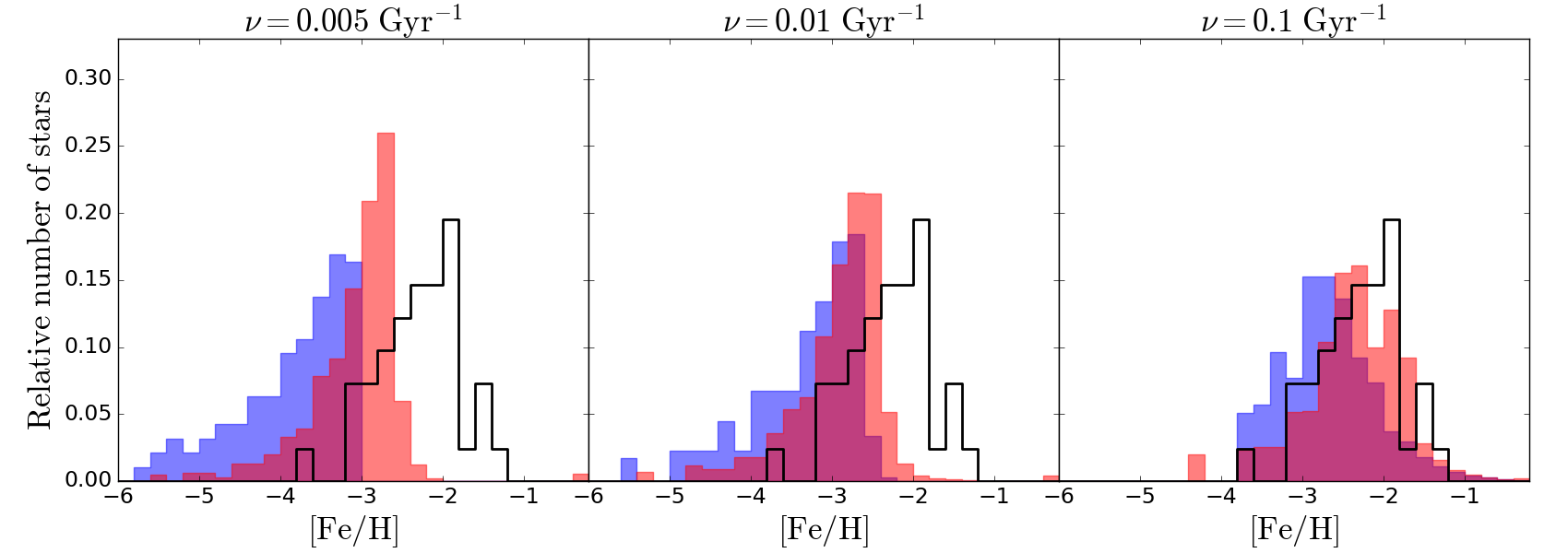}
  \caption{In the figure we reporte the observed MDF with the black line together with the predictions of the models; the results obtained with the Salpeter IMF are displayed in red, while in blue the ones with the IGIMF. For all the models the infall mass has been set to $M_{infall}=1.0\cdot 10^7\ \mathrm{M_{\odot}}$ while the SFE is changed. In the left panel we show the results for 1BooI, in the middle the ones for 2BooI and in the right panel the predictions for 3BooI.}
  \label{fig:mdfboo1}
\end{figure*}

The observed MDF has been built combining the high resolution data, used also for the analysis of the [$\alpha$/Fe] abundance ratios, with the low resolution ones, ending up with a dataset of 42 member stars.
In Figure \ref{fig:mdfboo1} we show the comparison between the observed MDF (black line) and the predicted ones adopting the Salpeter IMF (red) and the IGIMF (blue) for the models with $M_{infall}=1.0\cdot 10^7\ \mathrm{M_{\odot}}$. The models which best reproduce the observed distribution are the ones with $\nu=0.1\ \mathrm{Gyr^{-1}}$ (3BooI), especially the one adopting the Salpeter IMF. In all three panels the distributions obtained with the Salpeter IMF are peaked at higher [Fe/H] values than the IGIMF ones, as shown in Table \ref{tab:boo1res}. This is caused by the higher production of iron on short time-scales due to the higher number of CC SNe predicted by the Salpeter IMF (see Figure \ref{fig:sn2rateboo1}). The rapid decrease of the MDFs after the peak, visible in the first two panels, is due to the quench of the SF at 1 Gyr imposed by the observed SF history.

\subsubsection{Bo\"otes I: Summary}
 The model with the IGIMF which best reproduces the present-day stellar mass of Boo I is 1BooI, which is characterized by $M_{infall}=\ 1.0\cdot 10^7\ \mathrm{M_{\odot}}$ and $\nu=0.005\ \mathrm{Gyr^{-1}}$. Nevertheless, this model is not able to match the abundance ratios for any of the chemical elements analysed here. In addition, it predicts too many metal-poor stars and a lack of stars at higher [Fe/H], not fitting the observed MDF. On the contrary, the model 3BooI, with $\nu=0.1\ \mathrm{Gyr^{-1}}$, is able to reproduce the abundance ratios and the MDF but it does not predict the correct present-day stellar mass. 
Moreover, comparing the results obtained with both the IMFs, it emerges that the models adopting the Salpeter IMF better agree with data, in particular for the model with $M_{infall}=\ 2.5 \cdot 10^7\ \mathrm{M_{\odot}}$ and $\nu=0.01\ \mathrm{Gyr^{-1}}$ (5BooI), in accordance with what derived by \cite{vincenzo2014} for Boo I.

\subsection{Chemical evolution of Bo\"otes II}
\label{subsec:boo2}
Boo II has a present-day stellar mass that is one order of magnitude lower than Boo I, even though its mass of the dark matter halo is larger, as derived by \cite{koch2009}. They derived $M_{DM}= 3.3\cdot 10^6\mathrm{M_{\odot}}$ assuming the half-light radius estimated by \cite{martin2008} who found $r_L=51$ pc, the same values we adopted in our models. We assumed the fraction between the half-light radius and the core radius of the dark matter equal to $S=0.3$. 
For the infall mass we have assumed in our models the following three values: $M_{infall}= 2.5\cdot 10^5\ \mathrm{M_{\odot}}$, $M_{infall}= 5.0\cdot 10^5\ \mathrm{M_{\odot}}$ and $M_{infall}= 1.0\cdot 10^6\ \mathrm{M_{\odot}}$. The gas infall time-scale of this amount of gas has been set to $\tau_{infall}=0.005$ Gyr.   
Since a precise estimation of the star formation history of Boo II has never been done, we have assumed it to last 1 Gyr from the estimated average age of the galaxy determined by \cite{walsh2008}.
The other parameter that we have varied is the SFE; we have explored three values: $\nu=0.005\ \mathrm{Gyr^{-1}}$ ,$\nu=0.01\ \mathrm{Gyr^{-1}}$ and $\nu=0.1\ \mathrm{Gyr^{-1}}$. For the wind efficiency, instead, we have assumed $\omega=10 \ \mathrm{Gyr^{-1}}$ as for Boo I.

In Table \ref{tab:databoo2} we have summarized the input parameters of our chemical evolution models for Boo II. Finally, we have also varied the IMF adopting both the IGIMF and the Salpeter one.

\begin{table*}
\centering
\hspace*{-1.0 cm}
\caption{Input parameters used for all the chemical evolution models performed for Bo\"otes II. \textit{Columns:} (1) star formation efficiency, (2) wind efficiency, (3) infall time-scale, (4) star formation history, (5) total infall gas mass, (6) mass of the dark matter halo \citep{koch2009}, (7) half-light radius \citep{martin2008}, (8) ratio between the half-light radius and the dark matter effective radius, (9) initial mass function.}
\begin{tabular}{ccccccccc}
\hline
\hline
\multicolumn{9}{c}{\bf Bo\"otes II: parameters of the model} \\
\rule{0pt}{1.\normalbaselineskip}
$\nu$&$\omega$&$\tau_{inf}$&SFH&$M_{infall}$&$M_{DM}$&$r_L$&$S=\frac{r_L}{r_{DM}}$&IMF \\
$\mathrm{(Gyr^{-1})}$&$\mathrm{(Gyr^{-1})}$&$\mathrm{(Gyr)}$&$\mathrm{(Gyr)}$&$\mathrm{(M_{\odot})}$&$\mathrm{( M_{\odot})}$&$\mathrm{(pc)}$&\\[0.1cm] \hline
\noalign{\vskip 0.065in} 
0.005/0.01/0.1&10&0.005&$0-1$&$2.5/5.0/10.0\cdot 10^5 $&$3.3\cdot 10^6$&$51$&0.3&IGIMF/Salpeter\\
\hline
\hline
\end{tabular}

\label{tab:databoo2}
\end{table*}

 \begin{figure} 
 %\centering
 \hspace{-0.6cm}
  \includegraphics[width=1.05\columnwidth]{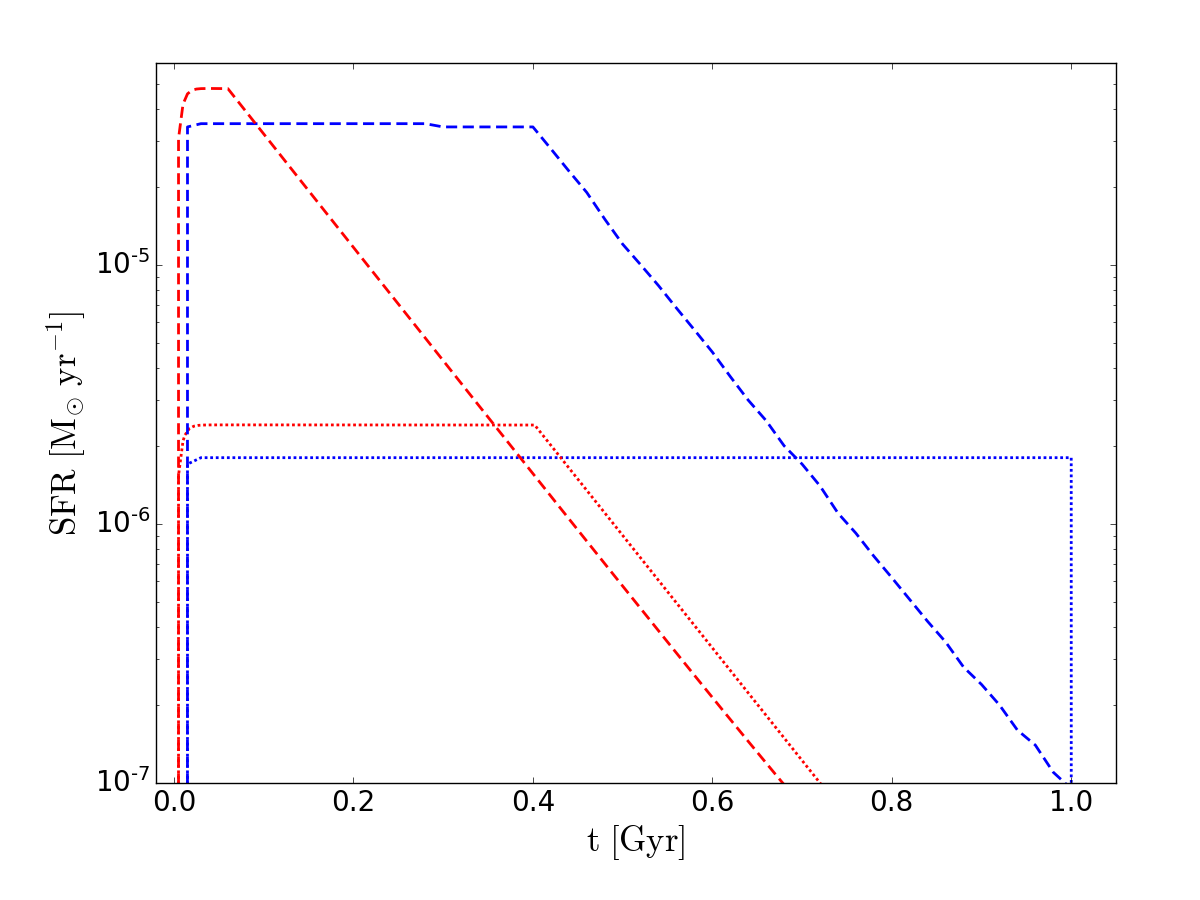}
  \caption{The star formation rate is shown as a function of time for $M_{infall}=5.0 \cdot 10^5\  \mathrm{M_{\odot}}$. The red lines represent the models with the Salpeter IMF while the blue lines refer to the results obtained with the IGIMF. With the dotted lines are shown the models for $\nu=0.005\ \mathrm{Gyr^{-1}}$ while with the dashed lines the ones for $\nu=0.1\ \mathrm{Gyr^{-1}}$. }
  \label{fig:sfrboo2}
\end{figure}

In Figure \ref{fig:sfrboo2} are shown the predictions for the SFR of Boo II. We have plotted here the results  obtained by assuming the Salpeter IMF in red, and the IGIMF in blue. All the models have been obtained supposing an infall mass of $M_{infall}=5.0 \cdot 10^5\  \mathrm{M_{\odot}}$ while the SFE has been varied. We have selected the two extreme values of $\nu=0.005\ \mathrm{Gyr^{-1}}$ and $\nu=0.1\ \mathrm{Gyr^{-1}}$. The graph is similar to the one obtained for Boo I presented in Figure \ref{fig:sfrboo1}. However, if we focus on the values of SFR reached here, we can see that they are quite lower than the ones we have obtained for Boo I. This is due to the lower infall mass we have assumed for Boo II. Furthermore, the differences between the Salpeter IMF and the IGIMF are enhanced here as a consequence of the higher  fraction of low mass stars predicted by the IGIMF, which do not release their gas in the ISM lowering the SFR. Finally, focusing on the time at which the SFR starts declining because of the onset of the galactic wind, we observe that in Boo II the wind is predicted to appear earlier for the models adopting the Salpeter IMF. This is due to the lower binding energy caused by the decrease of the mass of the dark matter halo and of the infall mass. On the contrary, the heavy truncation of the IGIMF causes a significant decrease of supernova explosions inducing a slower increase of the thermal energy of the gas which delays the onset of the galactic wind.

\begin{figure} 
 %\centering
 \hspace{-0.5cm}
  \includegraphics[width=1.05\columnwidth]{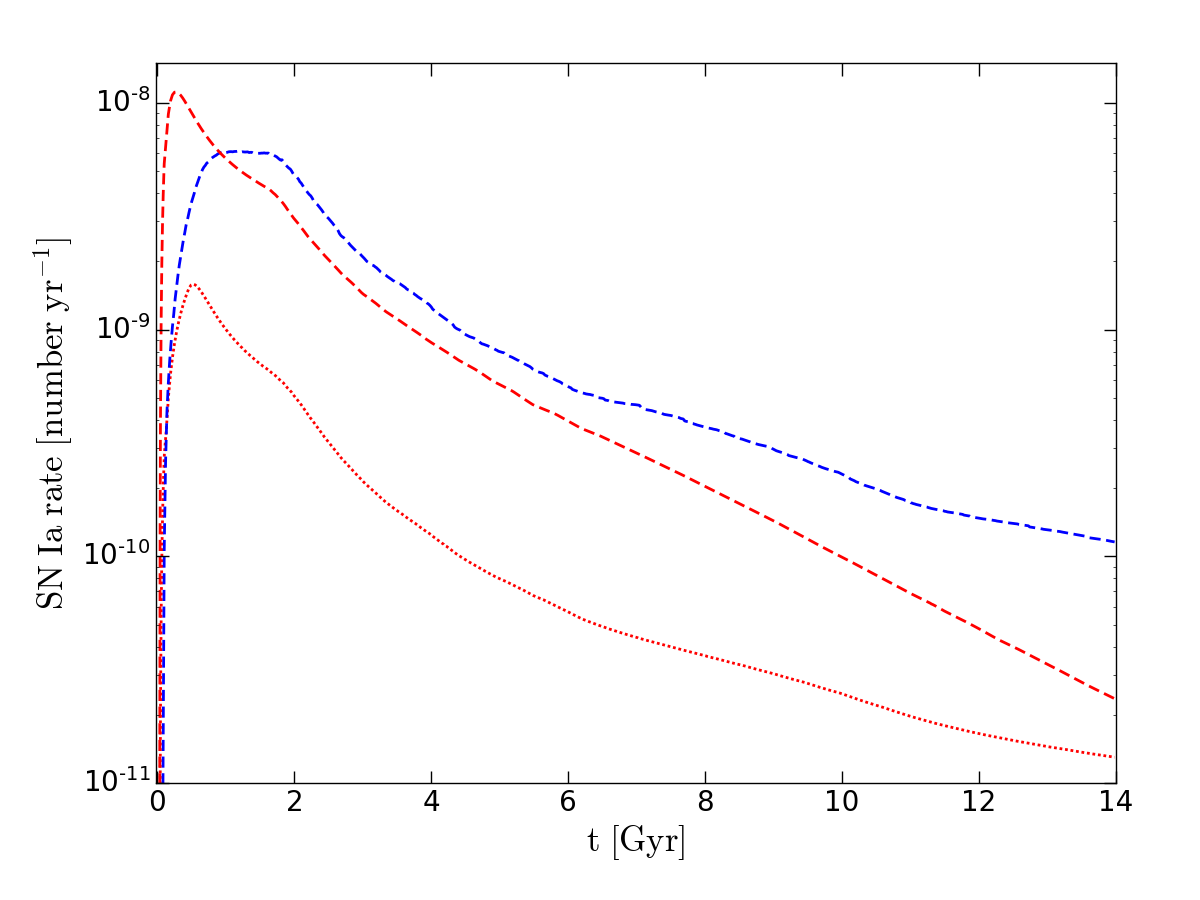}
  \caption{The Type Ia supernova rate is shown as a function of time for $M_{infall}=5.0 \cdot 10^5\  \mathrm{M_{\odot}}$. The red lines represent the models with the Salpeter IMF while the blue lines refer to the results obtained with the IGIMF. With the dotted lines are shown the models for $\nu=0.005\ \mathrm{Gyr^{-1}}$ while with the dashed lines the ones for $\nu=0.1\ \mathrm{Gyr^{-1}}$.}
  \label{fig:sne1arateboo2}
\end{figure}

In Figure \ref{fig:sne1arateboo2} is shown the rate of Type Ia SNe explosions for the same models of Figure \ref{fig:sfrboo2}. The model with $\nu=0.005\ \mathrm{Gyr^{-1}}$ is not plotted since it does not predict any Type Ia SN explosion. 
The explanation of the absence of these SNe can be understood looking at Figure \ref{igimf1}. What we infer is that for $\psi(t)<10^{-6} \mathrm{M_{\odot}yr^{-1}}$, the IGIMF does not predict the formation of stars with masses higher than $3\ \mathrm{M_{\odot}}$, thus no Type Ia SN progenitor is formed according to the model of SNe Ia adopted here ($\sim 3 \mathrm{M_{\odot}}$ is the minimum total mass of binary systems giving rise to SNe Ia). The heavy truncation of the IGIMF affects also the model with $\nu=0.1\ \mathrm{Gyr^{-1}}$ but at a lower extent. In fact, for such SFE, the SFR is high enough to permit the formation of binary systems more massive than $3\ \mathrm{M_{\odot}}$. However, only low binary systems could be formed which shifts the peak of the rate of SN explosions at later times if compared to the one predicted by the Salpeter IMF.

\begin{figure} 
 %\centering
 \hspace{-0.5cm}
  \includegraphics[width=1.05\columnwidth]{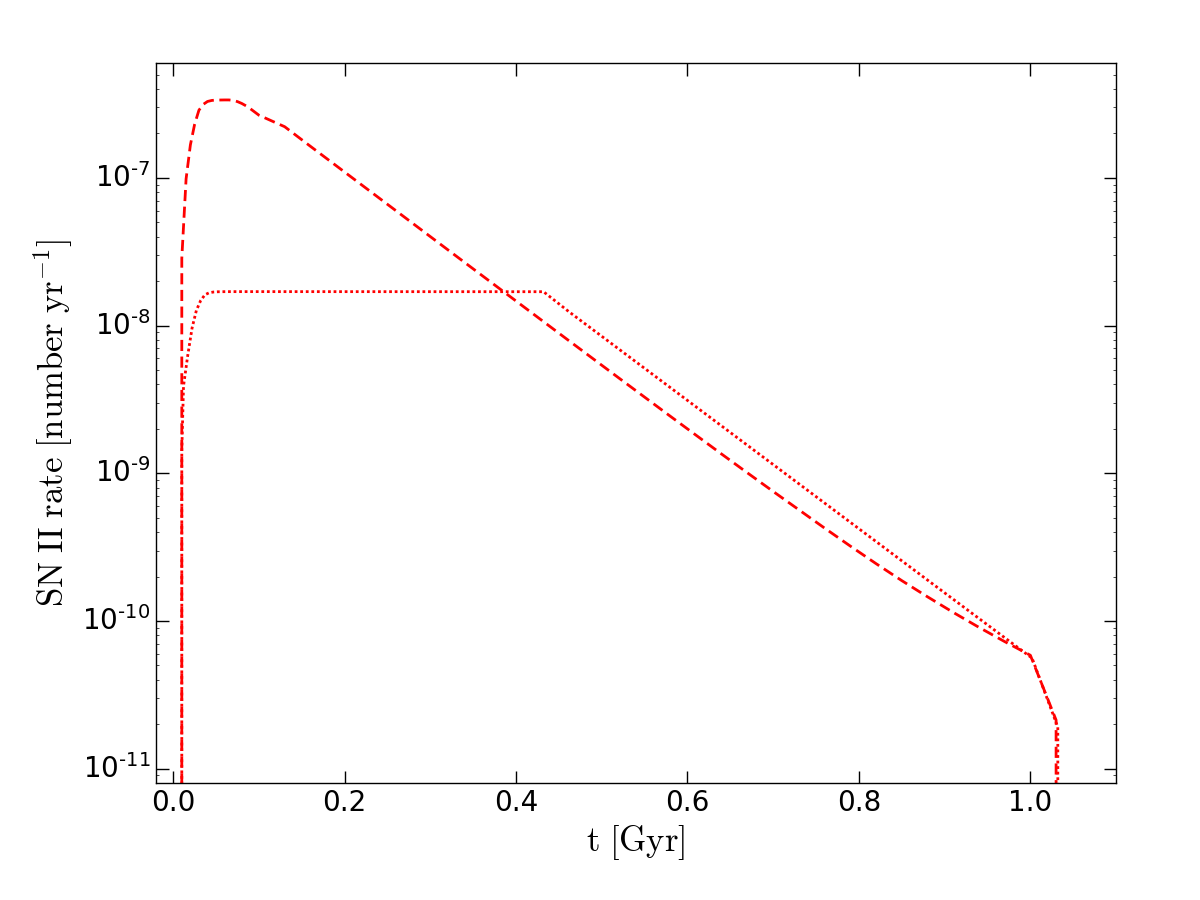}
  \caption{The Type II supernova rate is shown as a function of time for $M_{infall}=5.0 \cdot 10^5\  \mathrm{M_{\odot}}$. The red lines represent the models with the Salpeter IMF. With the dotted lines are shown the models for $\nu=0.005\ \mathrm{Gyr^{-1}}$ while with the dashed lines the ones for $\nu=0.1\ \mathrm{Gyr^{-1}}$. }
  \label{fig:sne2rateboo2}
\end{figure}

In Figure \ref{fig:sne2rateboo2} we show the predictions of the rate of Type II SN explosions obtained from the same models of Figure \ref{fig:sfrboo2} and \ref{fig:sne1arateboo2}. None of the two models adopting the IGIMF are here plotted since none of them are able to form stars more massive than $8\ \mathrm{M_{\odot}}$, the progenitors of Type II SNe. As in the case of Type Ia SNe, this is due to the strong truncation of the IGIMF for the SFR predicted for this systems. Comparing the rates of Type II SNe explosions predicted for Boo I (Figure \ref{fig:sn2rateboo1}) and the one for Boo II (Figure \ref{fig:sne2rateboo2}), we can see that the number of events per year are higher for Boo I because of its higher SFR.

 In Table \ref{tab:boo2res} we report the predictions of the chemical evolution models for Boo II. The second and the third column resume the input parameter we have varied while the other columns are devoted to the results we obtained for the present-day stellar mass, the time of the onset of the galactic wind and the [Fe/H] value at which the MDF reaches the peak. 

The present-day stellar mass estimated by \cite{martin2008} who derived $M_{\star}^{Salpeter}=\ 2.8^{+0.7}_{-0.5}\ \cdot 10^3\ \mathrm{M_{\odot}}$ and $M_{\star}^{Kroupa}=\ 1.4^{+1.3}_{-1.0}\ \cdot 10^3\ \mathrm{M_{\odot}}$ is well reproduced by the IGIMF models 4BooII and 5BooII. Both the predicted magnitudes derived for these models fit well the observed value of $-2.94^{+0.75}_{-0.74}$ mag. For the Salpeter IMF the best match is given by the model 8BooII. 

Most of the models adopting the IGIMF do not develop a galactic wind. In fact, these models predict a negligible number of core-collapse SNe and, in some cases, even Type Ia SNe, both responsible for heating up the ISM. As a consequence, the thermal energy of the gas predicted by these models does not increase in time, as in the case with SN explosions. Therefore the gas thermal energy never becomes larger than the gas binding energy and no wind develops.
In addition, looking at the two times of onset of the galactic wind for the models with $\nu=0.1\ \mathrm{Gyr^{-1}}$, the one derived for the lowest infall mass is higher than the others. The contrary is obtained with the Salpeter IMF and also with the IGIMF for more massive galaxies such as Boo I.  
Lowering the infall mass leads not only to a decrease of the binding energy of the gas but also to a drop of the SFR and consequently of the number of supernova explosions that heat up the ISM. For the most massive UFDs in our sample, this latter effect is less important than the former. However, diminishing the mass of the galaxy, such an effect becomes important, making the smallest systems experience the galactic wind at later times.

Given the negligible number of SN explosions predicted by the models with low SFE we have done some tests setting the wind efficiency $\omega$ proportional to the SN rates, as done by \citet{romano05}. The results do not differ from those obtained with a constant wind efficiency, since for the models such as 4BooII, 5BooII, in which a negligible number of Type II and Type Ia SNe are present, the wind never develops. In the case in which only Type II SNe are absent, the variation is also negligible.

\begin{table*}
\centering
%\hspace*{-2cm}
\caption{ The predictions of chemical evolution models for Bo\"otes II. The second and the third column contain the input parameters which have been varied in the models while the other six concern the model predictions. \textit{Columns:} (1) model name, (2) infall mass, (3) star formation efficiency, (4) present-day stellar mass derived with the IGIMF, (5) visual magnitude derived with the IGIMF,(6) present-day stellar mass derived with the Salpeter IMF, (7) time of the onset of the galactic wind assuming the IGIMF, (8) time of the onset of the galactic wind assuming the Salpeter IMF, (9) peak of the stellar MDF obtained with the IGIMF, (10) peak of the stellar MDF obtained with the Salpeter IMF.}

 \resizebox{1.85\columnwidth}{!}{
 \begin{tabular}{ccc|ccccccc}
\hline
\hline
\multicolumn{10}{c}{{\bf Bo\"otes II: model predictions}} \\[0.2cm]
\rule{0pt}{1.3\normalbaselineskip}

%Input parameters&&&Model Predictions\\
Model name&$M_{infall}$&$\nu$&$M_{\star,fin}^{IGIMF}$&$M_V^{IGIMF}$&$M_{\star,fin}^{Salpeter}$&$t_{wind}^{IGIMF}$&$t_{wind}^{Salpeter}$&$\mathrm{[Fe/H]_{peak}^{IGIMF}}$&$\mathrm{[Fe/H]_{peak}^{Salpeter}}$\\[0.2cm]
&($\mathrm{M_{\odot}} $)&($\mathrm{Gyr^{-1}} $)&($\mathrm{M_{\odot}} $)&(mag)&($\mathrm{M_{\odot}} $)&(Gyr)&(Gyr)&(dex)&(dex)\\[0.2cm] \hline   
\rule{0pt}{1.\normalbaselineskip}
\hspace{-0.15cm}
1BooII&$2.5 \cdot 10^{5}$&0.005&$7.2 \cdot 10^{2}$&&$3.1 \cdot 10^{2}$&No wind&0.37&&-3.1\\  
2BooII&$2.5 \cdot 10^{5}$&0.01&$1.4 \cdot 10^{3}$&&$4.4 \cdot 10^{2}$&No wind&0.25&&-2.9\\
3BooII&$2.5 \cdot 10^{5}$&0.1&$8.4 \cdot 10^{3}$&&$2.0 \cdot 10^{3}$&0.52& 0.06&-3.5&-2.3\\ \
&&&&&&&\\
4BooII&$5.0 \cdot 10^{5}$&0.005&$1.4 \cdot 10^{3}$&$-2.65$&$7.0 \cdot 10^{2}$&No wind&0.43&&-3.1\\  
5BooII&$5.0 \cdot 10^{5}$&0.01&$2.9 \cdot 10^{3}$&$-3.29$&$1.0 \cdot 10^{3}$&No wind&0.31&&-2.9\\
6BooII&$5.0\cdot 10^{5}$&0.1&$1.4 \cdot 10^{4}$&&$4.3 \cdot 10^{3}$&0.42& 0.07&-2.9&-2.5\\
&&&&&&&\\
7BooII&$1.0 \cdot 10^{6}$&0.005&$2.9 \cdot 10^{3}$&&$1.7 \cdot 10^{3}$&No wind&0.55&&-2.9\\  
8BooII&$1.0 \cdot 10^{6}$&0.01&$5.8 \cdot 10^{3}$&&$2.4 \cdot 10^{3}$&2.60&0.37&&-2.7\\
9BooII&$1.0\cdot 10^{6}$&0.1&$2.8 \cdot 10^{4}$&&$9.7 \cdot 10^{3}$&0.31& 0.09&-2.9&-2.5\\
\hline
\hline
\end{tabular}
}

\label{tab:boo2res}
\end{table*}

\begin{figure*}
        \centering
        \begin{subfigure}{1cm}
        \centering
        \hspace*{-1.4cm}
                    \includegraphics[width=0.4\linewidth]{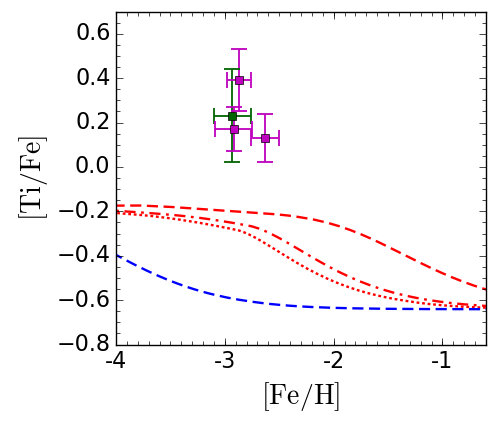}
               
  \label{fig:First_figure}
       \end{subfigure}%
%no, don't use ~!           ~ %add desired spacing between images, e. g. ~, \quad, \qquad etc.
      
    \hfill   
   
        \begin{subfigure}{6cm}
        \centering
        \hspace*{-1.8cm}
                    \includegraphics[width=0.85\linewidth]{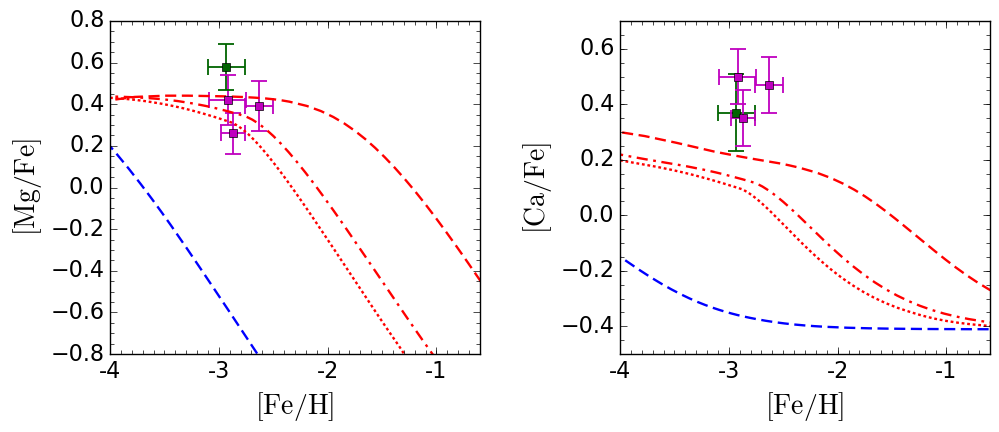}
                
  \label{fig:Second_figure}
       \end{subfigure}%
  
    \caption{In the figure we compare the $[\alpha$/Fe] versus [Fe/H] abundance ratios for Mg, Ti and Ca as observed in Bo\"otes II UFD member stars with the predictions of the chemical models with $\omega =10 \ \mathrm{Gyr^{-1}}$, $M_{infall}=5.0 \cdot 10^5\ \mathrm{M_{\odot}}$ for different values for $\nu$ and varying the IMF. In red we show the models with the Salpeter IMF while in blue the ones with the IGIMF. The models with $\nu=0.005\ \mathrm{Gyr^{-1}}$ (4BooII) are represented by the dotted line, the ones with $\nu=0.01\ \mathrm{Gyr^{-1}}$ (5BooII) by the dash-dotted line, while the models with $\nu=0.1\ \mathrm{Gyr^{-1}}$ (6BooII) are plotted with the dashed line. Concerning the data, the green squares refer to the data sample of \citet{koch2014}, the magenta ones are taken from \citet{ji2016}.}
   
      \label{fig:abundboo2}
 \end{figure*}

\subsubsection{Bo\"oets II: Adundance ratios and its interpretation}
The dataset consists of four RGB stars analyzed by \cite{koch2014} and \cite{ji2016}. The stars span a small iron abundance range with the most metal-poor one having [Fe/H] = $-2.93\pm0.17$ dex, while the most metal-rich has [Fe/H] = $-2.63\pm0.13$ dex. The abundances of Mg, Ti, and Ca are concentrated in an even smaller range with a maximum for Mg of about 0.3 dex. 
 
In Figure \ref{fig:abundboo2} we show the effects of changing the SFE and the IMF on the [$\alpha$/Fe] vs. [Fe/H] relations for the models assuming $M_{infall}=5.0 \cdot 10^5\  \mathrm{M_{\odot}}$. The two models adopting the IGIMF with the lowest SFEs do not show any iron enrichment remaining well below [Fe/H] = $-4.0$ dex. 
The knee appears in the model adopting the highest SFE but the decline starts at an extremely low [Fe/H] not matching the data. The models assuming the Salpeter IMF are able to fit the Mg abundances but underestimate the Ti and Ca ones. For Ti a possible explanation could be the uncertain yields for this element (see \citealp{romano2010}).

  \begin{figure*} 
 %\centering
 %\hspace{-0.3cm}
  \includegraphics[width=1.\textwidth]{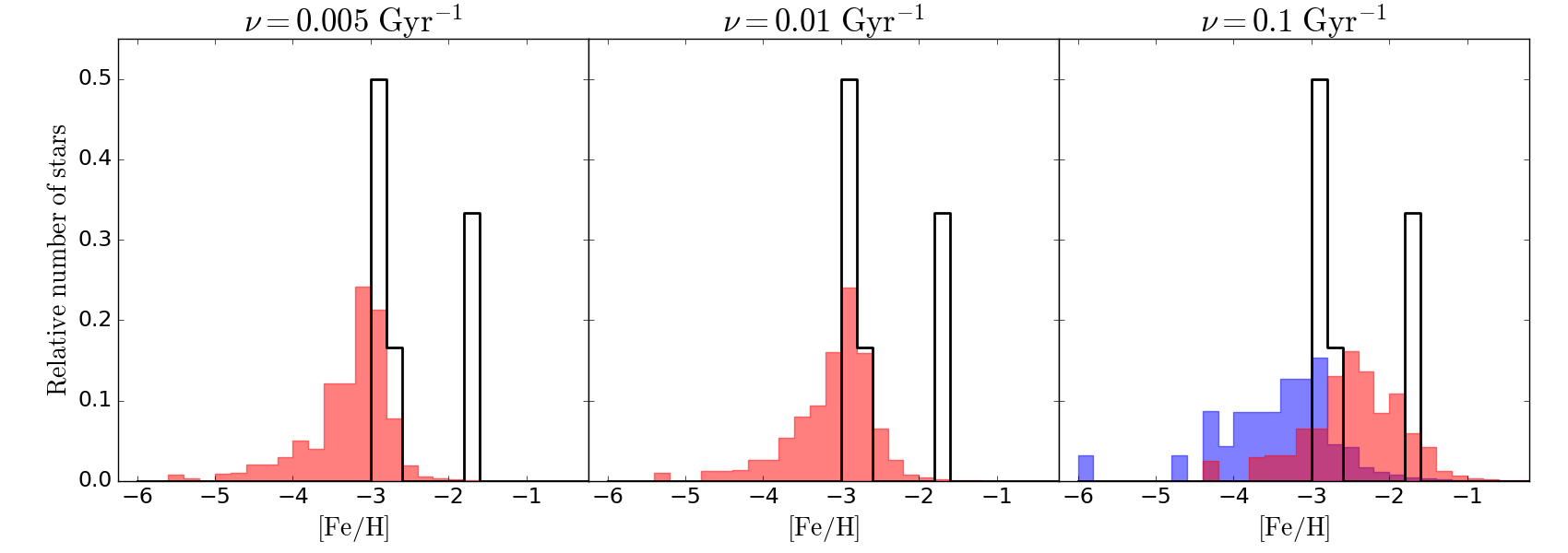}
  \caption{In the figure we report the observed MDF with the black line together with the predictions of the models; the results obtained with the Salpeter IMF are displayed in red, while in blue the ones with the IGIMF. For all the models the infall mass has been set to $M_{infall}=5.0\cdot 10^5\ \mathrm{M_{\odot}}$ while the SFE is changed. In the left panel we show the results for 4BooII, in the middle the ones for 5BooII and in the right panel the predictions for 6BooII.}
  \label{fig:mdfboo2}
\end{figure*}

\subsubsection{Bo\"otes II: MDF and its interpretation}
The dataset we have used here to built up the observed MDF is a very poor sample composed by six RGB member stars. Nevertheless some important conclusions can be drawn. 
In Figure \ref{fig:mdfboo2} we present the observed MDF (in black) together with the model predictions assuming $M_{infall}=5.0 \cdot 10^5\ \mathrm{M_{\odot}}$. In the first two panels, the MDFs obtained with the IGIMF are not plotted since, at the end of the star formation, the [Fe/H] is well below $-6.0$ dex. In the model 6BooII, instead, stars are formed above [Fe/H] = $-6.0$ dex, but they are still too metal-poor. An overabundance of stars at low [Fe/H] is predicted also by the models 4BooII and 5BooII with the Salpeter IMF. On the contrary, the model with $\nu=0.1\ \mathrm{Gyr^{-1}}$ (6BooII) reproduces quite well the observed MDF.
Comparing the observed and the predicted MDFs, the relative number of stars are very different but the reason of this discrepancy lies on the very poor sample we are dealing with.

\addtocontents{toc}{\protect\setcounter{tocdepth}{1}}
\subsubsection{Bo\"otes II: Summary}
\addtocontents{toc}{\protect\setcounter{tocdepth}{2}}

The models adopting the IGIMF are able to reproduce the present-day stellar mass if low SFEs are assumed. Nevertheless, the same models are not able to fit either the abundance ratios or the observed MDF. In almost all the cases we were not even able to plot them, since no enrichment has been predicted. For the models with the Salpeter IMF, instead, they agree quite well with the observational constraints, especially the model 8BooII characterized by $\nu=0.01\ \mathrm{Gyr^{-1}}$ and $M_{infall}=1.0\cdot 10^6\ \mathrm{M_{\odot}}$.

% Canes Venatici I Canes Venatici I Canes Venatici I Canes Venatici I Canes Venatici I Canes Venatici I Canes Venatici I Canes Venatici I Canes Venatici I Canes Venatici I Canes Venatici I Canes Venatici I Canes Venatici I

\subsection{Chemical evolution of Canes Venatici I}

\begin{table*}
\centering
\hspace*{-0.5 cm}
\caption{Input parameters used for all the chemical evolution models performed for Canes Venatici I. \textit{Columns:} (1) star formation efficiency, (2) wind efficiency, (3) infall time-scale, (4) star formation history \citep{weisz2014}, (5) total infall gas mass, (6) mass of the dark matter halo \citep{collins2014} obtained using \citet{martin2008} half-light radius values, (7) half-light radius \citep{martin2008}, (8) ratio between the half-light radius and the dark matter effective radius, (9) initial mass function.}
\begin{tabular}{ccccccccc}
\hline
\hline
\multicolumn{9}{c}{\bf Canes Venatici I: parameters of the model} \\
\rule{0pt}{1.\normalbaselineskip}
$\nu$&$\omega$&$\tau_{inf}$&SFH&$M_{infall}$&$M_{DM}$&$r_L$&$S=\frac{r_L}{r_{DM}}$&IMF \\
$\mathrm{(Gyr^{-1})}$&$\mathrm{(Gyr^{-1})}$&$\mathrm{(Gyr)}$&$\mathrm{(Gyr)}$&$\mathrm{( M_{\odot})}$&$\mathrm{( M_{\odot})}$&$\mathrm{(pc)}$&\\[0.1cm] \hline
\noalign{\vskip 0.065in} 
0.01/0.15&5/10&0.005/5&$0-6$&$2.5/2.7/3.5/4.5\cdot 10^7$&$1.9\cdot 10^7$&$564$&0.3&IGIMF/Salpeter\\
\hline
\hline
\end{tabular}

\label{tab:datacvn1}
\end{table*}

Canes Venatici I is the most massive and most extended galaxy in our sample, and it is sometimes considered more as a dSph rather than an UFD galaxy. The derived mass of its dark matter halo is $M_{DM}=1.9\cdot10^7\ \mathrm{M_{\odot}}$ \citep{collins2014}, with a half-light radius of $r_L=564$ pc \citep{martin2008}. As for the other galaxies, we have assumed the ratio between the  effective radius of the luminous (baryonic) component and the radius of the core of dark matter halo to be $S=0.3$. 
For the star formation history, we have supposed that it consisted in a single event lasted 6 Gyr as estimated by \cite{weisz2014}. 
The infall time-scale of such initial reservoir of gas has been assumed to be $\tau_{infall}=0.005$ Gyr, as for the other galaxies; however, with this value, we are not able to reproduce the cumulative SFH derived by \citet{weisz2014}. Therefore, we have run models with a longer $\tau_{infall}$ of $5$ Gyr. Concerning the SFE, it has been set to $\nu=0.15\ \mathrm{Gyr^{-1}}$ while the wind efficiency has been varied between $\omega=5$ and $10 \ \mathrm{Gyr^{-1}}$. 
In Table \ref{tab:datacvn1} are summarized the input parameters of our chemical evolution models. For every combination of SFE and infall mass we have ran two models, one adopts the Salpeter IMF and the other the IGIMF.

\begin{table*}
\centering
%\hspace*{-2cm}
\caption{ The predictions of chemical evolution models for Canes Venatici I. The second and the third column contain the input parameters which have been varied in the models while the other six concern the model predictions. \textit{Columns:} (1) model name, (2) infall mass, (3) star formation efficiency, (4) present-day stellar mass derived with the IGIMF, (5) visual magnitude derived with the IGIMF, (6) present-day stellar mass derived with the Salpeter IMF, (7) time of the onset of the galactic wind assuming the IGIMF, (8) time of the onset of the galactic wind assuming the Salpeter IMF, (9) peak of the stellar MDF obtained with the IGIMF, (10) peak of the stellar MDF obtained with the Salpeter IMF.}
 \resizebox{1.85\columnwidth}{!}{
 \begin{tabular}{ccc|cccccccc}
\hline
\hline
\multicolumn{11}{c}{{\bf Canes Venatici I: model predictions}} \\[0.2cm]
\rule{0pt}{1.3\normalbaselineskip}

%Input parameters&&&Model Predictions\\
Model name&$M_{infall}$&$\omega$&$\tau_{inf}$&$M_{\star,fin}^{IGIMF}$&$M_V^{IGIMF}$&$M_{\star,fin}^{Salpeter}$&$t_{wind}^{IGIMF}$&$t_{wind}^{Salpeter}$&$\mathrm{[Fe/H]_{peak}^{IGIMF}}$&$\mathrm{[Fe/H]_{peak}^{Salpeter}}$\\[0.2cm]
&($\mathrm{M_{\odot}} $)($\mathrm{Gyr^{-1}} $)&($\mathrm{Gyr}$)&($\mathrm{M_{\odot}} $)&(mag)&($\mathrm{M_{\odot}} $)&(Gyr)&(Gyr)&(dex)&(dex)\\[0.17cm] \hline   
\rule{0pt}{1.\normalbaselineskip}
\hspace{-0.15cm}
%1CVnI&$2.5 \cdot 10^{7}$&0.01&10&0.005&$1.5 \cdot 10^{5}$&&$7.7 \cdot 10^{4}$&0.78&0.49&-2.7&-2.7\\  
%1CVnI&$2.5 \cdot 10^{7}$&0.15&10&0.005&$6.3 \cdot 10^{5}$&$-8.76$&$3.6 \cdot 10^{5}$&0.15&0.09&-2.5&-2.3\\ 
I1CVnI&$2.5 \cdot 10^{7}$&10&0.005&$6.3 \cdot 10^{5}$&$-8.76$&&0.15&&-2.5&\\ 
S2CVnI&$3.5 \cdot 10^{7}$&10&0.005&&&$5.7 \cdot 10^{5}$&&0.11&&-2.1\\ 
%CVnI&$3.5 \cdot 10^{7}$&0.01&10&0.005&$2.3 \cdot 10^{5}$&&$1.3 \cdot 10^{5}$&0.83&0.58&-2.7&-2.5\\ 
%2CVnI&$3.5 \cdot 10^{7}$&0.15&10&0.005&$8.8\cdot 10^{4}$&&$5.7 \cdot 10^{5}$&0.15&0.11&-2.5&-2.1\\ 
I3CVnI&$2.7 \cdot 10^{7}$&5&5&$5.0 \cdot 10^{5}$&$-8.74$&&0.24&&-1.5&\\ 
S4CVnI&$4.5 \cdot 10^{7}$&5&5&&&$5.8 \cdot 10^{5}$&&0.05&&-1.3\\
%7CVnI&$5.0 \cdot 10^{7}$&0.01&10&0.005&$3.4 \cdot 10^{5}$&$-8.24$&$2.2 \cdot 10^{5}$&0.89&0.73&-2.5&-2.5\\
%8CVnI&$5.0 \cdot 10^{7}$&0.15&10&0.005&$1.4 \cdot 10^{6}$&&$9.3 \cdot 10^{5}$&0.17&0.16&-2.1&-1.9\\
\hline
\hline
\end{tabular}
}

\label{tab:cvn1res}
\end{table*}

  \begin{figure} 
 %\centering
 %\hspace{-0.6cm}
  \includegraphics[width=1.\columnwidth]{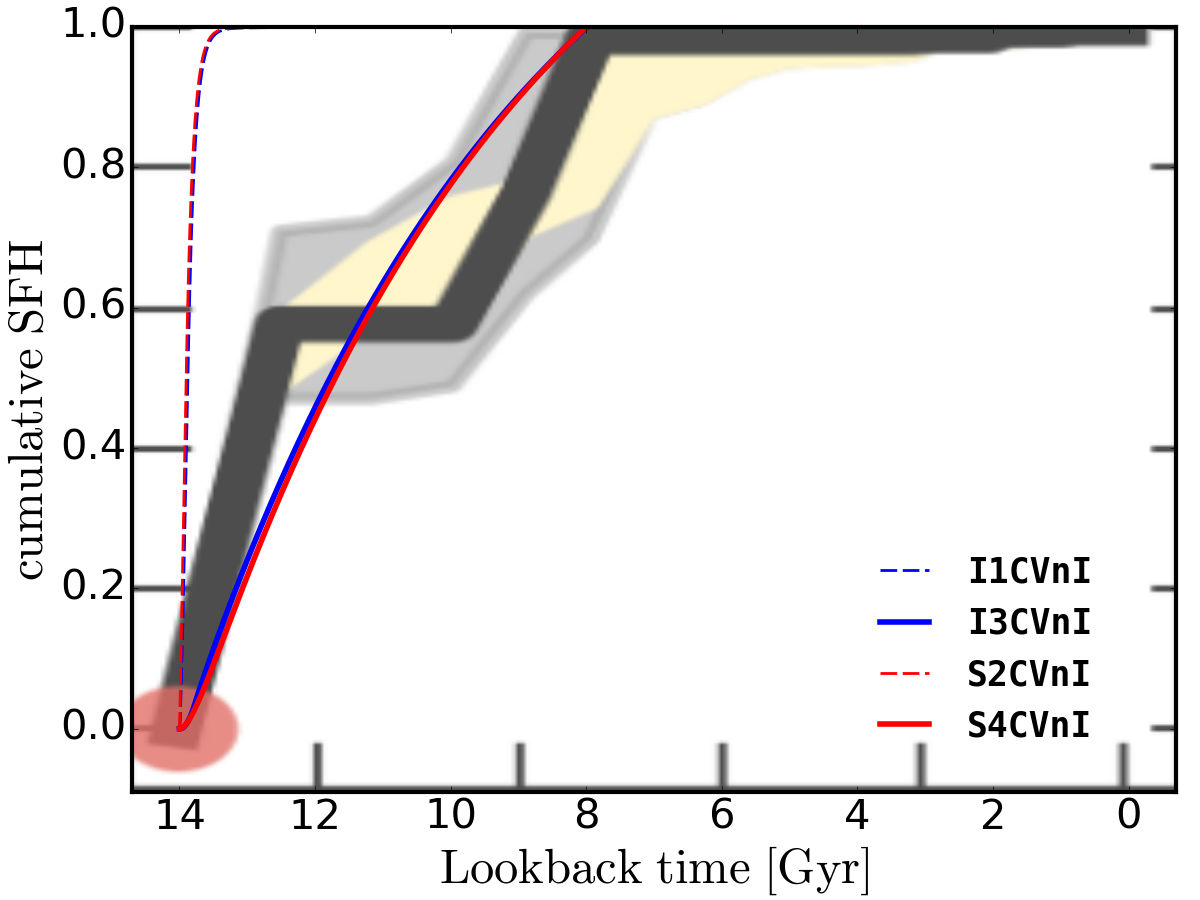}
  \caption{The cumulative star formation rate is shown as a function of time. The red lines represent the models with the Salpeter IMF while the blue lines refer to the results obtained with the IGIMF. The models with $\tau_{infall}=0.005$ Gyr are represented by the dashed lines. The solid and thick lines represent the two models with a longer infall timescale of 5 Gyr. In grey the total uncertainties (random and systematic) for the cumulative SFH taken from \citet{weisz2014} are shown.}
  \label{fig:cumsfr_cvn1}
\end{figure}

In Table \ref{tab:cvn1res} we have summarized the predictions of the models for CVn I which are able to reproduce the present-day stellar mass. The first letter in the name of models refers to the assumed IMF, 'S' for the Salpeter IMF and 'I' for the IGIMF.
From the second to the forth column the input parameters that have been varied through the models are listed, while the other columns are devoted to the results we have obtained using both the IMFs. Given the higher mass of the dark matter halo and the higher observed stellar mass of CVn I with respect to the other UFDs analyzed here, a higher $M_{infall}$ is needed to reproduce its present-day stellar mass. In fact, from \cite{martin2008} we have that $M_{\star}^{Salpeter}=\ (5.8\pm0.4)\ \cdot 10^5\ \mathrm{M_{\odot}}$ and $M_{\star}^{Kroupa}=\ (3.0\pm0.2)\ \cdot 10^5\ \mathrm{M_{\odot}}$. For the models with the IGIMF we have derived also the visual magnitude which can fit the observed one of $-8.73\pm0.06$ mag.

The predictions of the cumulative SFH for the aforementioned models are shown in Figure \ref{fig:cumsfr_cvn1}. It appears that only I3CVnI and S4CVnI can fit the observed trend derived by \citet{weisz2014}, given their very long infall timescale that prevents the steep increase of the cumulative SFH at early times.

\begin{figure*} 
 %\centering
% \hspace{-1.5cm}
  \includegraphics[width=0.85\textwidth]{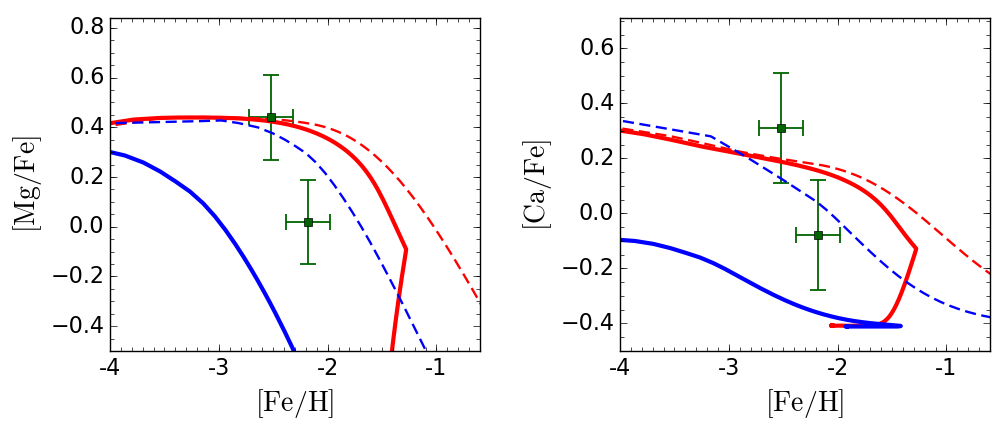}
  \caption{In the figure we compare the $[\alpha$/Fe] versus [Fe/H] abundance ratios for Mg and Ca as observed in CVn I UFD member stars with the predictions of the chemical models. In red we show the models with the Salpeter IMF while in blue the ones with the IGIMF. The models with $\tau_{infall}=0.005\ \mathrm{Gyr}$ are represented by the dotted lines while the ones with $\tau_{infall}=5\ \mathrm{Gyr}$ by the solid thick lines. The data have been taken from \citet{francois2016}.}
  \label{fig:abundcvn1}
\end{figure*}

\subsubsection{Canes Venatici I: abundance ratios and its interpretation}

The sample of stars with available high-resolution data for CVn I consists of only two stars analyzed by \cite{francois2016}. In Figure \ref{fig:abundcvn1} we show the observational data together with the predictions of the [$\alpha$/Fe] vs. [Fe/H] trends for the same models of Figure \ref{fig:cumsfr_cvn1}. What can be inferred is that only the model I1CVnI fits well the abundance ratios of the two stars, while the model I3CVnI is not able to reproduce the observations. Regarding the models adopting the Salpeter IMF, both of them predict the appearance of the knee at higher [Fe/H] values if compared with the models assuming the IGIMF; therefore, they are not able to fit the abundances of the star with lower [$\alpha$/Fe] ratios.  Moreover, in both the models with a longer infall timescale, for low [$\alpha$/Fe] ratios, the [Fe/H] value decreases. This behaviour is caused by the dilution of the enriched gas with the the gas of primordial composition that is still falling into the galaxy.
% three models fit quite well the abundance ratios of the two stars. Nevertheless, the two models adopting the Salpeter IMF do not match the observed present-day stellar mass. On the contrary, the model 3CVnI with the IGIMF is able to fit the data and to reproduce the mass in stars at the present time. The IGIMF models assuming a lower SFE, instead, are not able to match the observations because of their decrease in [$\alpha$/Fe] at very low [Fe/H]. 

  \begin{figure} 
 %\centering
 %\hspace{-0.6cm}
  \includegraphics[width=1.\columnwidth]{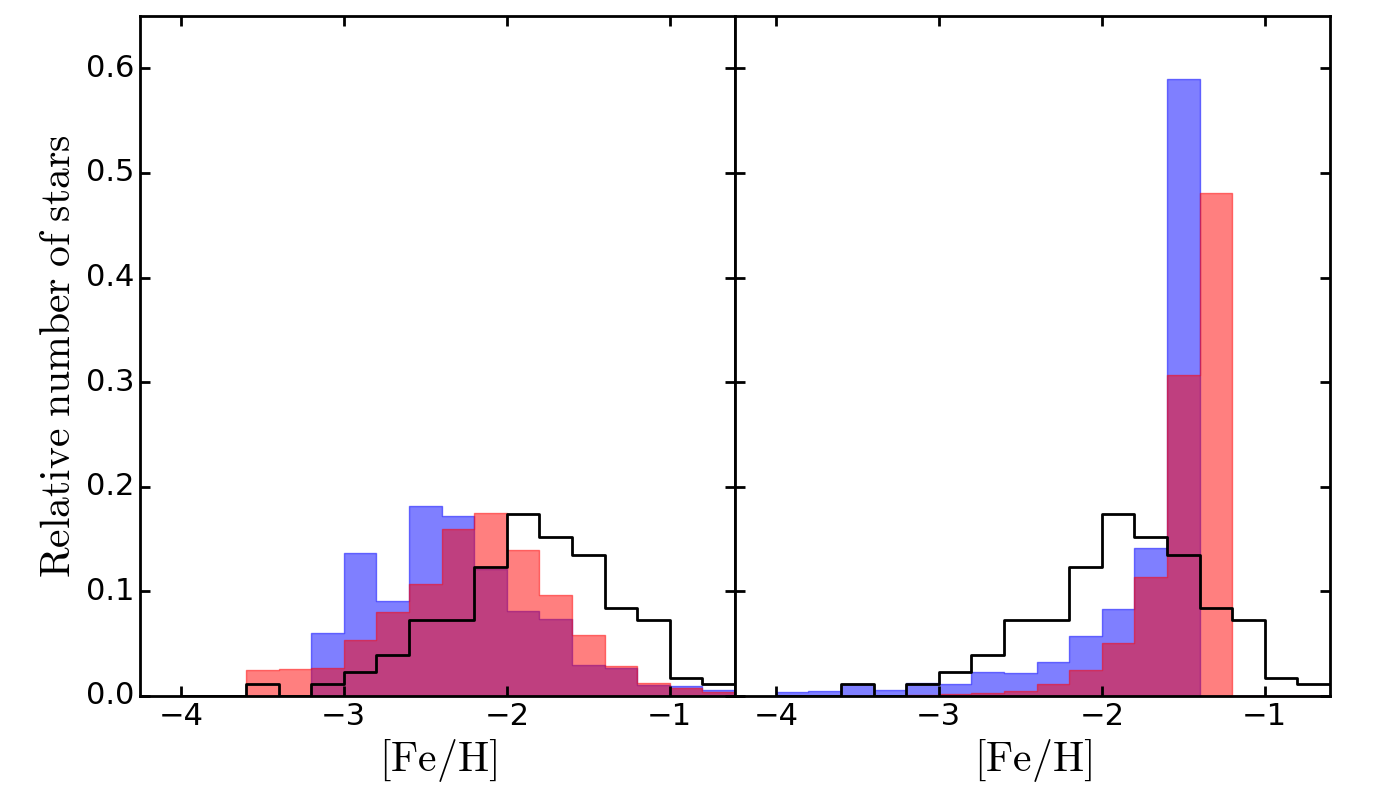}
  \caption{In the figure we report the observed MDF with the black line together with the predictions of the models;  the results obtained with the Salpeter IMF are displayed in red, while in blue the ones with the IGIMF. In the left panel we show the results for I1CVnI and S2CVnI while in the right panel the predictions for I3CVnI and S4CVnI are shown.}
  \label{fig:mdfcvn1}
\end{figure}

\subsubsection{Canes Venatici I: MDF and its interpretation}
The sample used to build up the observed MDF is composed by a rich statistical sample of 181 RGB stars analyzed by \cite{martin2007} and \cite{kirby2010} together with the two stars studied by \cite{francois2016}. In Figure \ref{fig:mdfcvn1} we present the predictions for the MDF in the case of long and short infall timescales.
The models assuming $\tau_{infall}=5$ Gyr predict a very high fraction of metal-rich stars at variance with observations. For short infall timescales, instead, too many metal-poor stars are predicted that are not detected by spectroscopic measurements. However, this mismatch could be influenced by the instrumentation capability to detect EMP stars because of their weak spectral lines.
As for the other two galaxies, the Salpeter IMF always predicts a lower number of metal-poor stars than the IGIMF, reproducing better the observations only if a short infall timescale is assumed.

% The models assuming low SFEs are not able to reproduce the observed MDF since all of them predict a peak of the MDF at too low [Fe/H] values. The models with $\nu=0.15\ \mathrm{Gyr^{-1}}$ best reproduce the distribution, especially for the Salpeter IMF. The models 4CVnI and 5CVnI are able to reproduce the present-day stellar mass but they cannot fit the observed MDF predicting too many metal-poor stars. 

\subsubsection{Canes Venatici I: Summary}

 In order to reproduce the observed present-day stellar mass of CVn I we need to assume a higher SFE and a higher infall mass with respect to the other UFDs. For this galaxy the models with a short infall timescale, typically associated to UFDs, do not reproduce the cumulative SFH derived from observations. In order to fit the SFH we have then increased $\tau_{infall}$. However, although these last models (I3CVnI and S4CVnI) are able to fit both the observed stellar mass and the cumulative SFH, they do not well reproduce the [$\alpha$/Fe] abundances, especially the model with the IGIMF. The MDF is also not well reproduced, since too few metal-poor stars are predicted. A better agreement with the observed abundances can be reached with model I1CVnI, in particular for the [$\alpha$/Fe] abundances, while the predicted MDF is peaked at a lower [Fe/H] if compared with the observed one (a sort of G-dwarf problem). Therefore, we cannot draw any firm conclusion on the basis of the present data. In the future, more data on abundance ratios will help to select the best model.

%In particular, we have derived that, for the IGIMF, the mass in stars today can be reproduced by the model 3CVnI, which assumes $M_{infall}=2.5 \cdot 10^7\ \mathrm{M_{\odot}}$ and $\nu=0.15\ \mathrm{Gyr^{-1}}$. Such a model can well reproduce the abundance ratios while it predicts too many metal-poor stars in comparison with the observed MDF (a sort of G-dwarf problem). Moreover, it can reproduce the present-day stellar mass even assuming the Salpeter IMF. However, if we adopt a Salpeter IMF the predicted decrease of the [$\alpha$/Fe] abundance appears at high [Fe/H], not matching the data. In addition, even if it fits better the MDF than the model assuming the IGIMF, the predicted number of metal-poor stars are higher than observed.
%Therefore, we can conclude that for CVn I it is not clear which IMF best fits the observational data. Nevertheless, both best models for the two adopted IMFs have been obtained with $\nu=0.15\ \mathrm{Gyr^{-1}}$ (model 3CVnI) a typical SFE of dSph galaxies, which is not surprising given the large mass (baryonic and dark matter) of this galaxy.

\subsection{The remaining galaxies}

 \begin{figure*}
 %centering
 %\hspace{-0.4cm}
  \includegraphics[width=0.85\textwidth]{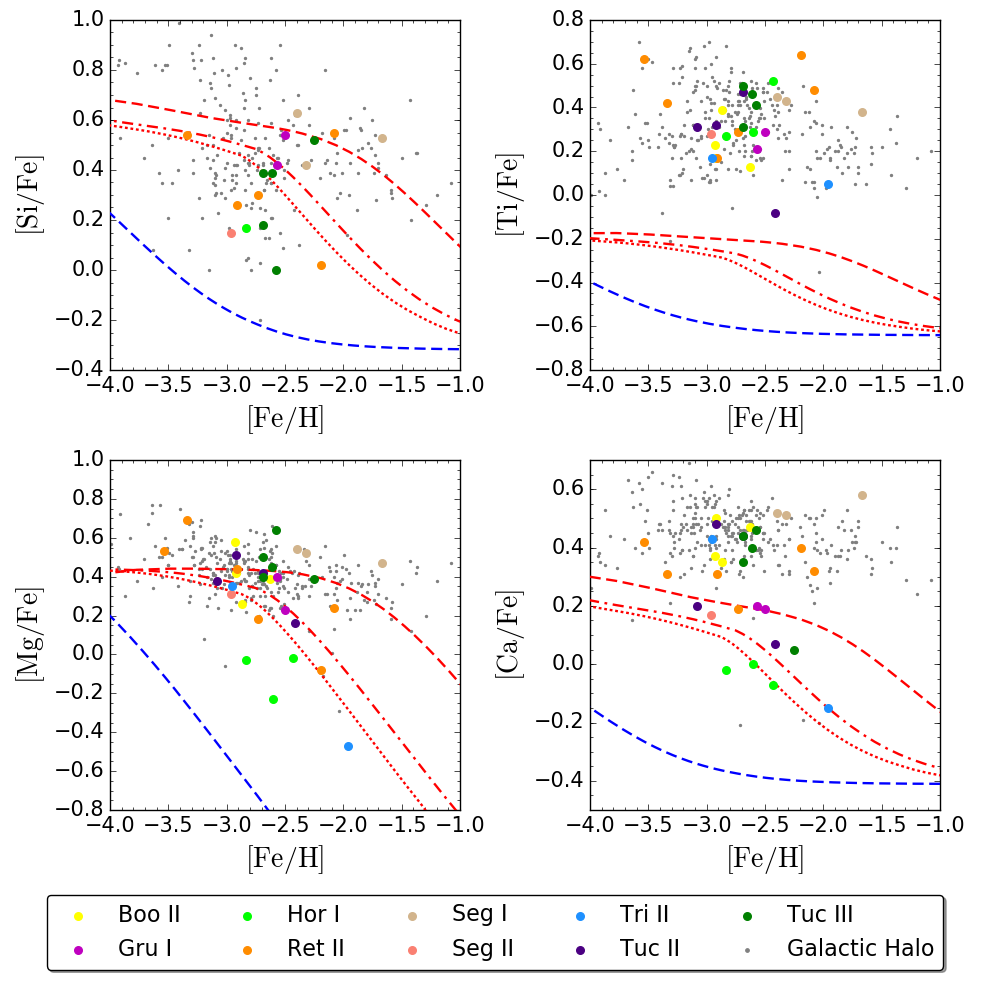}
  \caption{In the figure we compare the $[\alpha$/Fe] versus [Fe/H] abundance ratios for Si, Ti, Mg and Ca as observed in Boo II \citep{koch2014,ji2016}, Gru I \citep{ji2019}, Hor I \citep{nagasawa2018}, RetII \citep{ji2016b}, Seg I \citep{frebel2014}, Seg II \citep{roekir2014}, Tri II \citep{kirby2017,venn2017}, Tuc II \citep{chiti2018} and Tuc III \citep{hansen2017,marshall2018} UFDs member stars with the same model predictions of Figure \ref{fig:abundboo2}. The abundances of the Galactic halo stars have been taken from \citet{roederer2014}. The mean error on the UFD stars abundances are around $\sim 0.17$ dex for Fe while for $\alpha$-elements it is  around $\sim 0.25$ dex.}
  \label{fig:abbb2group}
\end{figure*}

Regarding Gru I, Hor I, Ret II, Seg I, Seg II, Tri II, Tuc II and Tuc III, the results resemble the ones of Boo II given the similar estimated stellar mass ($M_{\star}\sim 10^3\mathrm{M_{\odot}}$, for more details see \citealp{martin2008}, \citealp{bechtol2015} and \citealp{drlicawagner2015}). In Figure \ref{fig:abbb2group} we show the abundances of all the stars belonging to this subset of galaxies together with the model predictions obtained for Boo II in the [$\alpha$/Fe] vs. [Fe/H] plane. For all these small UFDs the IGIMF is not able to reproduce the observed abundances since it predicts a negligible number of SN explosions and consequently negligible $\alpha$ and iron enrichment.

 \begin{figure*} 
 %centering
 %\hspace{-0.4cm}
  \includegraphics[width=0.85\textwidth]{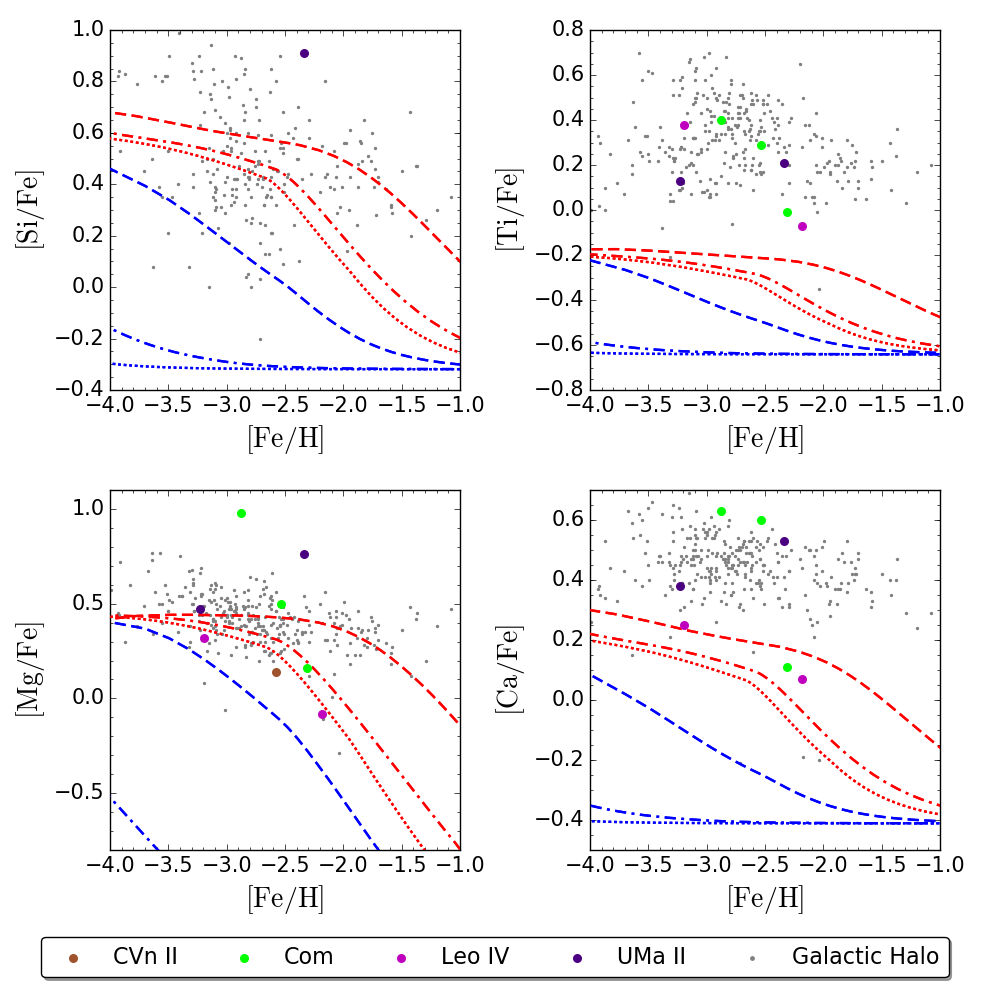}
  \caption{In the figure we compare the $[\alpha$/Fe] versus [Fe/H] abundance ratios for Si, Ti, Mg and Ca as observed in CVn II \citep{francois2016}, Com \citep{frebel2010}, Leo IV \citep{francois2016,simon2010} and UMa II \citep{frebel2010} UFDs member stars with the model predictions of CVn II whose input parameter are reported in Table \ref{tab:datacvn2}. The abundances of the Galactic halo stars have been taken from \citet{roederer2014}. The mean error on the UFD stars abundances are around $\sim 0.18$ dex both for Fe and $\alpha$-elements.}
  \label{fig:abbc2group}
\end{figure*}

Concerning the UFDs CVn II, Com, Leo IV and UMa II they have an observed mass in between the ones of Boo I and Boo II ($M_{\star}\sim 10^4\mathrm{M_{\odot}}$, \citealp{martin2008}). For these galaxies star formation, SNe Ia and II rates are similar to the ones of Boo I, but roughly an order of magnitude lower. This implies a lower production of $\alpha$ elements and iron from core-collapse SNe, before the appearance of the Type Ia ones and also a heavier truncation of the IGIMF. Consequently, the knee in the [$\alpha$/Fe] appears at lower [Fe/H] values than for Boo I, and therefore they do not fit the observational data, even assuming $\nu=0.1 \mathrm{ Gyr^{-1}}$. In Figure \ref{fig:abbc2group} we show the abundances for all the stars belonging in these four galaxies, together with the model predictions obtained for CVn II whose input parameters are summarized in Table \ref{tab:datacvn2}. Even for these galaxies the Salpeter IMF better fits the three observational constraints than the IGIMF. 

\begin{table*}
\centering
\hspace*{-1 cm}
\caption{Input parameters used for all the chemical evolution models performed for Canes Venatici II. \textit{Columns:} (1) star formation efficiency, (2) wind efficiency, (3) infall time-scale, (4) star formation history \citep{brown2014}, (5) total infall gas mass, (6) mass of the dark matter halo \citep{collins2014} obtained using \citet{martin2008} half-light radius values, (7) half-light radius \citep{martin2008}, (8) ratio between the half-light radius and the dark matter effective radius, (9) initial mass function.}
\begin{tabular}{ccccccccc}
\hline
\hline
\multicolumn{9}{c}{\bf Canes Venatici II: parameters of the model} \\
\rule{0pt}{1.\normalbaselineskip}
$\nu$&$\omega$&$\tau_{inf}$&SFH&$M_{infall}$&$M_{DM}$&$r_L$&$S=\frac{r_L}{r_{DM}}$&IMF \\
$\mathrm{(Gyr^{-1})}$&&$\mathrm{(Gyr)}$&$\mathrm{(Gyr)}$&$\mathrm{( M_{\odot})}$&$\mathrm{( M_{\odot})}$&$\mathrm{(pc)}$&\\[0.1cm] \hline
\noalign{\vskip 0.065in} 
0.005/0.01/0.1&10&0.005&$0-1$&$2.5\cdot 10^6$&$0.9\cdot 10^6$&$74$&0.3&IGIMF/Salpeter\\
\hline
\hline
\end{tabular}

\label{tab:datacvn2}
\end{table*}

\begin{figure*}
 \centering
 \hspace{-0.4cm}
  \includegraphics[width=0.85\textwidth]{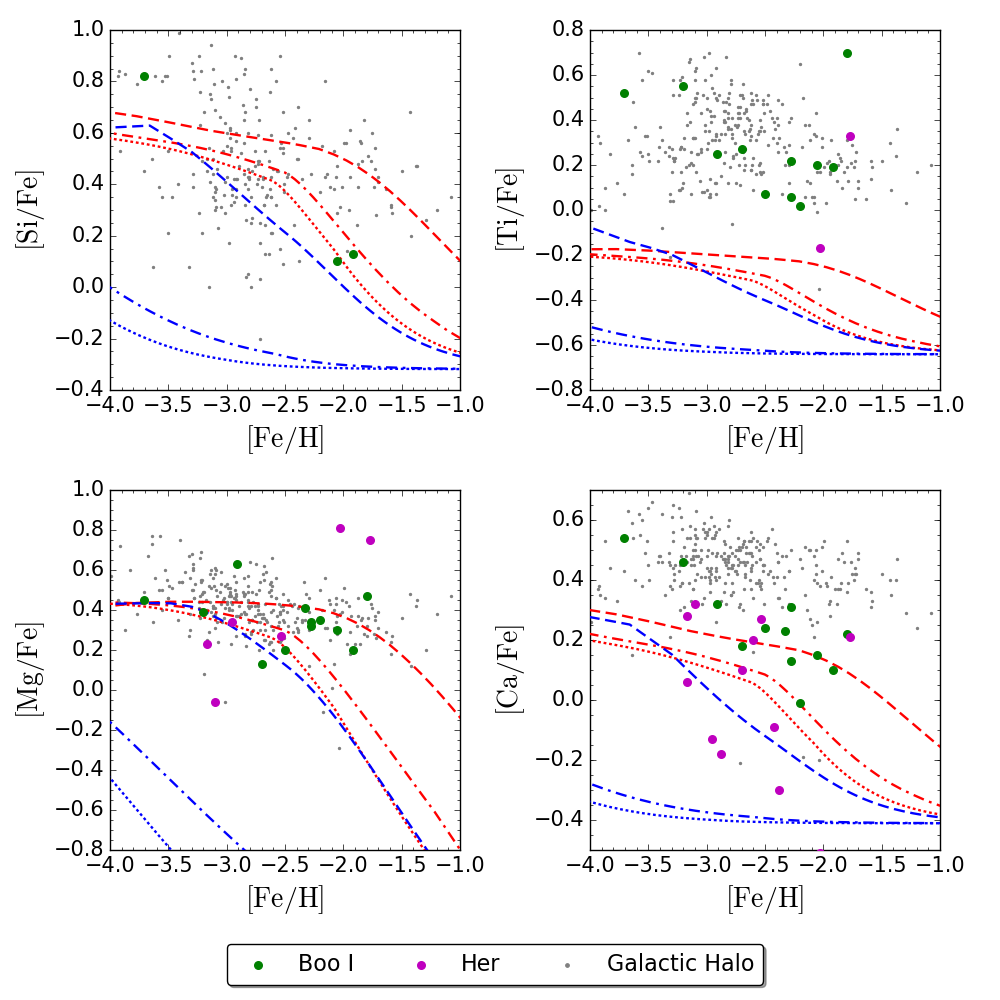}
  \caption{In the figure we compare the $[\alpha$/Fe] versus [Fe/H] abundance ratios for Si, Ti, Mg and Ca as observed in Boo I and Her UFDs member stars with the same model predictions of Figure \ref{fig:abundboo1}. For Her the abundances have been taken from \citet{koch2008}, \citet{aden2011} and \citet{francois2016} while the once of the Galactic halo stars from \citet{roederer2014}. The mean error on the UFD stars abundances are around $\sim 0.18$ dex both for Fe and $\alpha$-elements.}
  \label{fig:abbb1group}
\end{figure*}

The last galaxy we have analyzed is Her whose physical features are quite similar to Boo I ones. We have preferred to show the results for Boo I, since for this galaxy more stars have been studied and a larger sample of $\alpha$-element abundances is available. In Figure \ref{fig:abbb1group} are shown the abundances of Her and Boo I together with the model predictions for Boo I galaxy. Focusing on calcium, the model with the IGIMF assuming $\nu=0.1 \mathrm{Gyr^{-1}}$ reproduces quite well the observed abundances of Hercules. However, as obtained for Boo I, the final mass derived for this model is more than twice the observed value, making the IGIMF a worse IMF parametrization than the Salpeter one.

In Figures \ref{fig:abbb2group}, \ref{fig:abbc2group} and \ref{fig:abbb1group} we have also plotted the abundances of the Galactic halo stars in order to compare them with the UFD ones. It emerges that, unlike UFD stars, the Galactic halo ones do not display any decline in the [$\alpha$/Fe] ratio caused by the appearance of Type Ia SNe. The discrepancy is particularly clear for Boo I, Her, Hor I, Ret II, Tri II, Tuc II.
The overlap at extremely low [Fe/H] is not relevant since, at these metallicities, the enhancement in $\alpha$-elements is a common feature to systems at the beginning of their star formation history. Better discriminating factors are the $s$-process elements such as Ba \citep{spitoni2016} given the different ratios with iron at low metallicities in systems with low SFR.
Therefore, due to the particular behaviour of the [$\alpha$/Fe] ratio as discussed before and the lack of data for UFDs we cannot draw firmer conclusion.

%regarding the galaxies that do not show the decline in [$\alpha$/Fe], the possible explanations are:

%\textbf{
%\begin{itemize}
%\item they have been the building blocks of the Galactic halo even though \citet[two-infall+outflow]{brusadin2013} suggest a very high halo SF efficiency $\nu=2 \mathrm{Gyr^{-1}}$ completely different from the typical UFD value;
%\item they have undergone a very short star formation history, concluded before the appearance of SNe Ia as suggested for Segue I given its high $\alpha$ abundances at quite high metallicities \citep{frebel2014};
%\item the paucity of stars with available abundance data. Stars with low [$\alpha$/Fe] at high [Fe/H] could not have been detected and studied yet. 
%\end{itemize} 

\section{Conclusions}
\label{sec:conclusions}

In this work we have modeled the chemical enrichment history of sixteen UFD galaxies focusing on the results obtained for three of them, taken as prototypes of the least and the most massive UFDs in our sample. The novelty of this work consists in the adoption of a more physical IMF called IGIMF, in addition to the canonical Salpeter IMF to test whether it is able to better reproduce the observational constraints of these galaxies. 
We have adopted an updated version of the numerical code of \cite{lanfranchiematteucci2004}, \cite{vincenzo2014} and \cite{vincenzo2015}. The adopted model takes into account gas infall and outflow, detailed stellar nucleosynthesis for SNe core-collapse, SNe Ia and AGB stars and different prescriptions for the IMF.

Given their very low observed stellar mass, UFDs should have been characterized by a small initial reservoir of gas accreted on short time-scales. In our model we impose to reproduce the observed present-day stellar mass and the cumulative SFH of each galaxy, while our unknowns are the gas and its chemical composition. In order to reproduce the estimated present-day stellar masses in presence of galactic winds, we have supposed that the most massive UFDs have been formed by the accretion of an infall mass with primordial composition of $M_{infall}\sim 10^7 \ \mathrm{M_{\odot}}$ while, for the least massive ones, of $M_{infall}\sim 10^5 \ \mathrm{M_{\odot}}$. The gas falls at a rate obeying a decaying exponential law with an infall time-scale equal to $\tau_{infall}=0.005\ \mathrm{Gyr}$, with the exception of CVn I where models with a longer infall timescale have been tested. The stellar masses are usually derived from the SED (spectral energy distribution) method by adopting an IMF. In the cases adopting the IGIMF, since no mass determinations are available, we reproduced the observed visual magnitude by means of a spectro-photometric model \citep{vincenzo2016}. 
%\item In almost all the galaxies we have analyzed the [$\alpha$/Fe] abundance ratios are observed to decline at very low [Fe/H] which suggests that the efficiency of star formation in these systems is very low ($\nu=0.005-0.01  \ \mathrm{Gyr^{-1}}$). The consequent low SFRs, in fact, causes a poor enrichment of the ISM in iron by core-collapse SNe  before the appearance of Type Ia ones.

In light of that, we summarize here the main results and conclusions we have obtained in this work:

\begin{enumerate}

\item We considered detailed stellar feedback and included galactic wind in the models. We concluded that the best agreement with data has been achieved assuming a wind efficiency (mass loading factor) $\omega=10 \ \mathrm{Gyr^{-1}}$ as it was also derived by \citet{vincenzo2014}, \citet{lanfranchiematteucci2004} and \citet{romano2019} for UFDs. It is worth noting that for the smallest galaxies the galactic wind never occurs for low SFE due to the small or negligible number of SNe. 

\item With the very low SFRs predicted for our 16 galaxies ($10^{-4}-10^{-6}  \ \mathrm{M_{\odot}yr^{-1}}$), the metallicity-dependent IGIMF proposed by \citetalias{recchi} strongly suppresses the formation of massive stars as well as decreases substantially the production of SNe Ia progenitors. For the most massive galaxies it implies a negligible production of core-collapse SNe at low SFEs, while for the least massive ones the same result is obtained also for higher SFEs ($\nu= 0.1  \ \mathrm{Gyr^{-1}}$). The consequence of this strong truncation of the IGIMF leads to a negligible enrichment in iron and $\alpha$-elements.

%{\bf
%\item The poor chemical enrichment due to the low SFEs, necessary to reproduce the observed stellar mass, does not allow us to fit the observational data. The discrepancy is visible either in the [$\alpha$/Fe] vs. [Fe/H] or in the MDF diagrams.
%}

 %\item Comparing the IGIMF with the Salpeter IMF predictions for the same input parameters, we found that, at the low SFRs typical of UFDs, the IGIMF predicts: 
% \begin{itemize}
% \item[$-$] higher present-day stellar masses due to the higher number of low mass stars which die on long time-scales;
% \item[$-$] lower [Fe/H] values at which the decline of the [$\alpha$/Fe] abundance ratio appears since less core-collapse SN progenitors are supposed to form;
% \item[$-$] lower [Fe/H] values at which the MDF reaches its peak due to the lower number of massive stars enriching the gas at early times.
% 
% \end{itemize}
% 

 \item The models with the IGIMF best fitting the present-day stellar mass and the cumulative SFH are not able to match the other two observational constraints, underestimating the [$\alpha$/Fe] ratios at the [Fe/H] values of the analyzed stars and producing too many extremely metal-poor stars than observed. These discrepancies are the more evident the lower is the observed stellar mass of the galaxy, because of the low chemical enrichment predicted by the more truncated IGIMF. Therefore, the IGIMF does not seem to work in the regions of very low SFR and metallicity typical of UFDs.

\item In the case of CVn I, which is considered more as a dSph galaxy, a short infall timescale, typical of UFDs, does not allow us to reproduce the cumulative SFH derived by observations, since it predicts a too step rise of the cumulative SFH. However, if a longer infall timescale is assumed, the MDF and the [$\alpha$/Fe] abundances are not well reproduced, for both the two IMFs. Therefore we are not able to determine which IMF better reproduces the four observational constraints of this galaxy. More data on this galaxy are necessary before drawing firm conclusions.

\item We are forced to conclude that the models with the Salpeter IMF are able to better reproduce the data than the IGIMF, at least for the version adopted here (\citetalias{recchi}). Perhaps an IGIMF with a weaker dependence on the SFR and a stronger dependence on metallicity would better fit the properties of these extremely small galaxies. Recently, a new IGIMF has been proposed by \citet{yan2017} and \citet{jerabkova2018} which enhances the dependence of the slopes of IMF on metallicity, extending it to the low mass range. Such IMF has been applied to the chemical evolution of elliptical galaxies by \citet{yan2019}. 

Nevertheless, the models assuming the Salpeter IMF that we have selected as the ones best reproducing the present-day stellar mass match well the [$\alpha$/Fe] trends but, for a large number of galaxies, they predict too many metal-poor stars than what have been derived by observations. This is similar to the G-dwarf problem but, with the exception of CVnI which has a longer SF episode, it persists even for longer infall time-scales than the one adopted here ($\tau_{inf}>0.005$ Gyr). The explanation of such a discrepancy could be an observational bias: the samples we have used to build up the MDFs could be influenced by the instrumentation capability to detect extremely low metallicity stars because of the weak spectral lines characterizing these stars.

\item Comparing the abundances of UFD stars with the Galactic ones, we suggest that at least a fraction of UFDs could not be the building blocks of the halo given the different [$\alpha$/Fe] trends observed for UFDs and the Galactic halo. To draw firmer conclusions more data about UFDs are necessary and more elements should be studied such as barium \citep{spitoni2016}.

%the evidence of a decline in [$\alpha$/Fe] not observed for halo stars.

\end{enumerate}

For future projects it will be interesting to put more stringent constraints on the stellar mass and IMF of UFDs by using photo-chemical evolution models that reproduce at the same time the observed chemical abundances and photometric properties (including the CMD) of UFD stars, taking into account the intrinsic incompleteness of observations (see also \citealp{vincenzo2016}). Finally, another promising follow-up will be to study the neutron-capture chemical abundances in UFDs with our chemical-evolution model including different IMFs and nucleosynthetic scenarios like merging neutron stars (see also \citealp{matteucci2014, vincenzo2015}).

\section*{Acknowledgements}
We thank the anonymous referee for his/her useful suggestions.
F. V. acknowledges support from the European Research Council Consolidator Grant funding scheme (project ASTEROCHRONOMETRY, G.A. n. 772293) and the support of a Fellowship from the Center for Cosmology and AstroParticle Physics at The Ohio State University. 

%%%%%%%%%%%%%%%%%%%%%%%%%%%%%%%%%%%%%%%%%%%%%%%%%%



\bibliographystyle{mnras}
\bibliography{paper_ufd} % if your bibtex file is called example.bib

%%%%%%%%%%%%%%%%%%%%%%%%%%%%%%%%%%%%%%%%%%%%%%%%%%

%%%%%%%%%%%%%%%%%% APPENDICES %%%%%%%%%%%%%%%%%%%%%
%
%\appendix

%%%%%%%%%%%%%%%%%%%%%%%%%%%%%%%%%%%%%%%%%%%%%%%%%%

% Don't change these lines
\bsp	% typesetting comment
\label{lastpage}
\end{document}